%% file: submit.tex
\documentclass[aps,showpacs,a4paper,floatfix,twocolumn,prc,amsmath,amssymb,nofootinbib]{revtex4-1}

\usepackage{xcolor,ulem}

\usepackage[utf8]{inputenc}
\usepackage[T1]{fontenc}
\usepackage[english]{babel}

\usepackage{lmodern}

\usepackage{booktabs} 
\usepackage{array} 
\usepackage{multirow} 
\usepackage{tabularx} 
\usepackage{subcaption} 
\usepackage{graphicx}
\usepackage{subcaption} 
\usepackage{amsmath}
\usepackage{amssymb}
\usepackage{mathtools}
\usepackage{bm} 
\usepackage{graphicx}
\usepackage{epsfig,hyperref}
\usepackage{color} 
\usepackage{morefloats}
\usepackage{rotating}
\usepackage{relsize}
\usepackage{url}

\usepackage{ragged2e}

\usepackage{booktabs}
\usepackage{soul}
\usepackage{array}

\usepackage{csquotes}

\begin{document}

\title{
Medium effects on light clusters from heavy-ion collisions within a relativistic mean-field description
}

\author{Tiago Cust{\'o}dio$^1$}
\author{Francesca Gulminelli$^2$}
\author{Alex Rebillard-Souli{\'e}$^2$}
\author{Diego Gruyer$^2$}
\author{R{\'e}mi Bougault$^2$}
\author{Tuhin Malik$^1$}
\author{Helena Pais$^1$}
\author{Constan\c ca Provid{\^e}ncia$^1$}

\affiliation{$^1$CFisUC, Department of Physics, University of Coimbra,
  3004-516 Coimbra, Portugal. \\
$^2$Normandie Univ., ENSICAEN, UNICAEN, CNRS/IN2P3, LPC Caen, F-14000 Caen, France. }


\date{\today}
\begin{abstract}

Central $^{136,124}$Xe$+^{124,112}$Sn collisions from INDRA data are
analysed using a Bayesian inference on 
light nuclei multiplicities to estimate the thermodynamical parameters and in-medium modification of the cluster self-energies within a relativistic mean-field model. 
An excellent description of experimentally measured abundances of H and He isotopes is obtained. 
We examine two possible modelling of in-medium effects as an increased in-medium 
effective mass,
or an increased vector repulsion. We show that these physical pictures cannot be discriminated by the data. In both cases, the temperature dependence of the meson couplings 
leads to a faster weakening of the light cluster abundances with temperature than previous studies predicted.  Possible  
systematic errors due to 
out-of-equilibrium effects affecting the experimental 
abundances, are considered by repeating the Bayesian inference 
with reduced information.
The abundance prediction 
of the species excluded from the constraint
is well compatible with the experimental data, suggesting that there is no a priori need of accounting for non-equilibrium effects or finite state interactions that potentially affect the deuteron yield.

\end{abstract}

\maketitle

\newpage

\section{Introduction}

Light nuclear clusters may be found in extreme astrophysical events 
such as in supernovae (SN) \cite{Arcones2008,Sumiyoshi:2008qv,Furusawa:2013tta} and binary neutron star mergers (BNS) \cite{Bauswein:2013yna,Rosswog2015,Radice:2018pdn,Navo:2022xle,Psaltis:2023jvk}.
The knowledge of their abundances is 
an important ingredient of the so-called general purpose equations of state (EoS) of warm and low-density nuclear matter \cite{Oertel2017}, essential for modeling the dynamical evolution of these extreme astrophysical events. 

Even though there are 
some ab initio calculations for clusterized nuclear matter in a dense medium \cite{Ropke:2011tr,Ropke2015,Ropke_2020,Ren_2024}, these approaches cannot be applied to the extremely large range of proton fractions, temperatures, and densities explored by SN and BNS.
Therefore, phenomenological nuclear models 
are mostly used 
for the construction of general purpose EoS \cite{CompOse}. In particular, relativistic mean-field models with embedded Lorentz covariance have been proposed,  treating nuclear clusters as independent quasiparticles, where the interaction with the surrounding medium is guaranteed through their coupling to mesonic fields \cite{Typel2009,Pais_2019}. Being phenomenological by nature, these couplings need to be calibrated on experimental 
data.

One of the first attempts to calibrate the in-medium modifications of cluster properties through heavy-ion collisions 
was based on the extraction of chemical 
constants, assuming ideal Boltzmann distributions for the light particles yields 
in ref. \cite{Qin2011}.

A similar analysis was later performed on a different data set 
in Refs.\cite{Pais2020,Pais2020prl}, where 
a parametrized correction was added to the Boltzmann 
expression in an attempt to account for in-medium corrections. The resulting 
chemical equilibrium constants 
were compared to 
relativistic mean-field (RMF) 
predictions in order to 
calibrate the cluster-meson couplings. 

The studies published in  Refs.\cite{Pais2020,Pais2020prl} have only considered the equilibrium constants to calibrate the models used and did not analyze  the experimental particle multiplicities.
In Ref.\cite{Custodio_prl_2025}, this approach was shown to fail in reproducing the experimentally measured mass fractions of $^2$H, $^3$H, $^3$He, while only partially reproducing $^4$He abundances.

In Ref. \cite{Custodio_prl_2025}, a step was taken towards improving the theoretical description. 
The modified ideal gas assumption was dropped and the experimentally inaccessible baryonic density and temperature were taken, together with the cluster coupling, as the unknown parameters to be inferred by a Bayesian analysis 
performed on the directly measured particle mass fractions.
This resulted in an excellent description of cluster abundances.  The determined scalar cluster couplings exhibit an explicit temperature dependence.

In this work, we 
outline the steps that led to the  Bayesian analysis conducted in Ref.\cite{Custodio_prl_2025} in greater detail. We also investigate the possible degeneracy between the cluster-$\sigma$ and $\omega$-meson couplings \cite{Ferreira:2012ha}, and consider whether this affects the results obtained in Ref. \cite{Custodio_prl_2025}. 
We will also consider whether detection issues or out-of-equilibrium effects could impact the results of Ref. \cite{Custodio_prl_2025}.  In particular, we will focus on the deuteron, since its low binding energy means that deuterons could undergo final-state interactions that do not respect the detailed balance implied by the equilibrium hypothesis.

The paper is organized as follows: in Section \ref{section_formalism}, we introduce
the RMF model and the Bayesian inference methods; Section \ref{section:data_analysis} 
details the data analysis methodology and explores the degeneracy between cluster-$\sigma$-meson and cluster-$\omega$-meson couplings; in Section \ref{section:deuterons} possible deuteron out-of-equilibrium effects and detection problems are considered. 
Finally, in Section \ref{section:conclusions}, some conclusions are drawn.

\section{Formalism}\label{section_formalism}

\subsection{Relativistic mean-field model}\label{section:rmf_formalism}

To model the thermodynamic equilibrium in our system, we use a relativistic mean-field formalism that includes nucleons and light nuclear clusters as independent quasi-particles and takes into account the expected in-medium effects of light clusters in a dense medium \cite{Typel2009,Ferreira:2012ha,Pais:2015xoa,Pais2018}.   
 The interactions 
 are mediated by the exchange of virtual mesons: the isoscalar-scalar $\sigma$-meson, the isoscalar-vector $\omega$-meson and  the isovector-vector $\rho$-meson. 
The corresponding Lagrangian density reads \cite{Pais2018}
\begin{equation}\label{Lagrangian_nlm_inicial}
	\mathcal{L}=\sum_{i=n,p} \mathcal{L}_{i} + \sum_{\substack{j={}^{2}\text{H},{}^{3}\text{H},\\{}^{3}\text{He},{}^{4}\text{He}}} \mathcal{L}_{j} +\mathcal{L}_m + \mathcal{L}_{nl} ~.
\end{equation} 

The first term describes the nucleons ($i=n,p$)
\begin{eqnarray}
	\mathcal{L}_i= \bar{\Psi}_i & \left[ i \gamma_{\mu} \partial^{\mu} - m_i + g_{\sigma N} \sigma - g_{\omega N} \gamma_{\mu} \omega^{\mu} \right. \\ & \left. - g_{\rho N} \gamma_{\mu} \vec{I}_i\cdot \vec{\rho^{\mu}}  \right]  \Psi_i ~, \nonumber
\end{eqnarray}
where $\Psi_i$ is the nucleon Dirac field and $\vec{I}_i$ is the isospin operator. The
quantities $g_{mN}$ ($m=\sigma, \omega, \rho$) are the meson-nucleon couplings, which are fitted to reproduce nuclear properties. In this work, we present results for two very different RMF parametrizations. The first one, the FSU parameterization \cite{Todd-Rutel:2005yzo}, belongs to the set of nonlinear RMF models for which the meson-nucleon couplings are constant. On the other hand, the DD2 parameterization \cite{Typel2009} is an example of an approach that considers density dependent meson-nucleon couplings \cite{Typel2009}. Both parametrizations have been calibrated to the ground state properties of finite nuclei and, in the case of FSU, also to their linear response, making them suitable for the description of low-density matter. The difference between the two predictions can be taken as an estimation of the model dependence of our results.

The second term of Eq.(\ref{Lagrangian_nlm_inicial}) defines the light clusters Lagrangian density, including the interaction with the mesons. For fermionic clusters with spin $1/2$ (${}^{3}\text{H}$, ${}^{3}\text{He}$) it reads
\begin{equation}\label{key}
	\mathcal{L}_{j}=\bar{\Psi}_{j}\left[\gamma_{\mu}i D_{j}^{\mu} - M_{j}^* \right]\Psi_{j}, ~ j={}^{3}\text{H},{}^{3}\text{He} ~,
\end{equation} 
whereas for bosonic clusters with spin 0 (${}^{4}\text{He}$) and 1 (${}^{2}\text{H}$), the Lagrangian density is given by, respectively,
\begin{eqnarray}
	\mathcal{L}_j(S=0)&=& \frac{1}{2}\left( iD_{j}^{\mu} \Psi_{j}
  \right)^*\left( iD_{\mu j} \Psi_{j}
  \right)     -
 \frac{1}{2}\Psi_{j}^*(M_j^*)^2\Psi_{j}, \nonumber
  \\ && \phantom{}   j={}^{4}\text{He} ~,   \\ 
		\mathcal{L}_j(S=1)&=&\frac{1}{4}\left( iD_{j}^{\mu}
      \Psi_{j}^{\nu}-iD_{j}^{\nu}
      \Psi_{j}^{\mu} \right)^*\left(
      iD_{\mu j} \Psi_{\nu j}-iD_{\nu
      j} \Psi_{\mu j} \right)  \nonumber \\ &&
            -\frac{1}{2}\Psi_{j}^{\mu*}(M_j^*)^2\Psi_{\mu j},\qquad j={}^{2}\text{H} ~, 
\end{eqnarray}
with 
\begin{equation}\label{key}
	i D_{j}^{\mu}=i \partial^{\mu} - g_{\omega j} \omega^{\mu} - g_{\rho j}\vec{I}_{j}\cdot \vec{\rho^{\mu}} ~.
      \end{equation}

The meson-cluster couplings are defined as $g_{\sigma j}=x_{s} A_j g_{\sigma N}$, 
$g_{\omega j}=x_{\omega} A_j\,g_{\omega N}$, where $A_j$ is the mass number of cluster and the 
coupling ratios $x_{s}(\rho,T)$, $x_{\omega}(\rho,T)$ measure the  in-medium modification of the cluster self-energies, and are considered as unknown parameters to be inferred from 
the experimental data. The consistence of the analysis requires that the same coupling should describe different data sets, but a density and/or temperature dependence may result from the inference.
Some degeneracy is expected between the couplings of the $\sigma$ and $\omega$ mesons to the clusters, as discussed in \cite{Ferreira:2012ha}. In the present study, the coupling to the $\omega$ meson is first fixed ($x_{\omega}=1$) and the coupling to the $\sigma$ is  calibrated to the experimental data. Later in Section \ref{section:xw_calibration}, we will redo some of the calculations, fixing  the coupling $x_{s}=1$ and inferring the $x_{\omega}$ 
coupling  
instead. 

 The effective mass of the cluster $j$, $ M_j^* $, is given by 
\begin{equation}\label{effective_mass_clusters}
	M_j^*=A_j m - g_{\sigma j}\sigma - (B_j^0 +\delta B_j) ~,
\end{equation}
where $m$ is the nucleon mass, $ B_j^0 $ is the tabulated vacuum
binding energy of the light cluster $ j $, and $ \delta B_j $ is a
binding energy shift that is included to take Pauli blocking effects in an
  effective way, as first introduced in Ref.\cite{Pais2018}.
It is defined as
\begin{equation}\label{eq:binding_energy_shift}
	\delta B_j=\frac{Z_j}{\rho_0}(\epsilon_p^*-m \rho_p^*)+ \frac{N_j}{\rho_0}(\epsilon_n^*-m \rho_n^*) ~,
\end{equation}
where $ Z_j $, $ N_j $ are the number of protons and neutrons, respectively. The gas energy density $\epsilon_j^*$ and nucleonic density $ \rho_j^* $ are given by
\begin{eqnarray}
	\epsilon_j^*&=&\frac{1}{\pi^2}\int_0^{p_{F_j}(\rm gas)} p^2 e_j(p) (f_{j+}(p)+f_{j-}(p)) dp \label{enerstar_clusters} \\
	\rho_j^* &=&\frac{1}{\pi^2}\int_{0}^{p_{F_j}(\rm gas)}  p^2 (f_{j+}(p)+f_{j-}(p)) dp \label{rhostar_clusters} ~,
\end{eqnarray}
where $p_{F_j}(\text{gas})=(3 \pi^2 \rho_j)^{1/3} $ is the Fermi momentum of nucleon $j$ defined using the zero temperature relation between the density and the Fermi momentum, $f_{j\pm}$ are the usual Fermi distribution functions for the particles and anti-particles, and $e_j=\sqrt{p_j^2+m^{*2}}$ is
the corresponding single-particle energy of the nucleon $j$.

The Lagrangian density of the meson fields $\mathcal{L}_{m}$ in Eq.(\ref{Lagrangian_nlm_inicial}) is given by 
\begin{eqnarray}
\mathcal{L}_{m}&=&\frac{1}{2}\partial_{\mu}\sigma\partial^{\mu}\sigma - \frac{1}{2}m_{\sigma}^2 \sigma^2 \nonumber \\ 
 && - \frac{1}{4}\Omega^{\mu\nu}\Omega_{\mu\nu} + \frac{1}{2}m_{\omega}^2\omega_{\mu}\omega^{\mu} \nonumber \\	
	&& - \frac{1}{4}\vec{R}^{\mu\nu} \cdot \vec{R}_{\mu\nu} + \frac{1}{2}m_{\rho}^2\vec{\rho}_{\mu} \cdot \vec{\rho}^{\mu}	~,
\end{eqnarray} 
with $ \Omega_{\mu\nu}=\partial_{\mu} \omega_{\nu}-\partial_{\nu}
\omega_{\mu} $ and $ \vec{R}_{\mu\nu}=\partial_{\mu}
\vec{\rho}_{\nu}-\partial_{\nu} \vec{\rho}_{\mu} +g_\rho(\vec{\rho}_\mu \times \vec{\rho}_\nu) $.

Finally, the last term of Eq.(\ref{Lagrangian_nlm_inicial}) includes non-linear meson terms to take into account the impact of the variation of the density in the nuclear interaction, controlling several EoS properties such as the compression modulus, the effective nucleon mass, the EoS softness and nuclear symmetry energy, see \cite{Pais2018,Custodio2020}:
\begin{eqnarray}
      \mathcal{L}_{nl} & = &  - \frac{\kappa}{3!}g_{\sigma N}^3\sigma^3 - \frac{\lambda}{4!}g_{\sigma N}^4 \sigma^4  + \frac{\zeta}{4!}g_{\omega N}^4(\omega_{\mu} \omega^{\mu})^2  \nonumber \\ && + \Lambda_{\omega} g_{\rho N}^2 \vec{\rho}_{\mu} \cdot \vec{\rho}^{\mu} g_{\omega N}^2 \omega_{\mu}\omega^{\mu} ~. 
 \end{eqnarray}
These terms are only present in models that do not have density dependent couplings, e.g. FSU. 

For density dependent models such as DD2, the density dependence of the nuclear interaction is taken into account through density dependent nucleon couplings such as:
\begin{eqnarray}\label{key}
		g_{iN}(\rho_B)&=&g_{iN}(\rho_0) a_i\frac{1+b_i (x+d_i)^2}{1 + c_i(x+d_i)^2}, \hspace{0.2cm} \text{$i= \sigma $, $ \omega $} \label{sigma_omega_density_dependence}\\ [10pt]
		g_{i N}(\rho_B)&=&g_{i N}(\rho_0)\text{exp}[-a_{\rho}(x-1)] , \hspace{0.2cm} \text{$ i=\rho $} \label{rho_density_dependence}
	\end{eqnarray}
	where the parameters $ a_i $, $ b_i $, $ c_i $, $  d_i $ are specific to each density dependent RMF model (given in \cite{Typel2009} for DD2), $ \rho_0 $ is the symmetric nuclear saturation density (which is also model dependent), and $ x=\rho_B/\rho_0 $. The cluster-meson couplings depend on the density through the nucleon couplings.

\subsection{Data sampling} \label{sec:samples}
For all analysis in this work,  we use  INDRA data sets corresponding to the center of mass (CM) emitting source of free nucleons and light nuclei produced with four different entrance channels $^{136,124}$Xe$+^{124,112}$Sn at 32 MeV/nucleon \cite{Bougault_2018}. To find
statistical ensembles composed of nucleons and light clusters, 
only central events (the most violent) are considered followed by an angular selection to reduce secondary decays from sources other than the intermediate velocity \footnote{ The intermediate velocity region being defined between the projectile-like and the target-like velocities} one. 
Intermediate energy heavy-ion collisions 
(10-100 MeV/nucleon) 
are a strongly dynamical process, and even in the most central collisions global chemical equilibrium might not be achieved. A weaker assumption consists in considering the dynamics of the expansion by sorting the data in bins of the average Coulomb-corrected particle velocities $v_{\text{surf}}$ in the CM frame \cite{Qin2011,indra}.
\\
This variable being correlated to the dynamics of the expansion, this extra sorting is more likely to produce samples which are statistically distributed, such that an effective sample dependent  temperature of the source can be defined \cite{Qin2011}. An indication of the statistical character of the samples was indeed experimentally obtained for all Hydrogen and Helium isotopes with the exception of ${}^{6}\text{He}$, from an isoscaling analysis~\cite{Rebillard-Soulie:2023rqu}.
 The deviations observed for ${}^{6}\text{He}$ were tentatively attributed to finite-size effects in Ref.~\cite{Rebillard-Soulie:2023rqu}, and this isotope is therefore excluded from all the analyses in the present work. 
The four different entrance channels and 13 different velocity bins leaves us with 52 data sets potentially associated with different sets of the control parameters $(\rho,T,y_p)$.
The global proton fraction $y_p$ of the samples can be safely deduced from the data with minimal hypotheses on the undetected neutrons \cite{indra}, 
but the baryonic density $\rho$ and the temperature $T$ are not measured experimentally and must be obtained indirectly through the comparison to the theoretical model described in the previous Section. The possible model dependence of the results will be quantified by comparing the results obtained with the two different DD2 and FSU models.

\subsection{Bayesian Inference Framework}\label{section:bayesian_framework}

In this work, a Bayesian inference on experimental particle mass fractions $\omega_{AZ}$ 
is performed to estimate the unmeasured theoretical parameters. These latter, collectively noted as $\theta=\{\rho,T,x\}$, include both the 
thermodynamic conditions of the system (baryonic densities and temperatures) and 
the scalar cluster coupling ratio $x_s$ ($x_{\omega}$ in Section \ref{section:xw_calibration}). To this aim, for each given experimental sample built as described in Section \ref{sec:samples}, the Bayes' theorem reads:
\begin{equation} \label{eq:Bayes}
    p \left( {\bf\theta}| \{\omega_{AZ} \}\right )=\frac{p_{{\bf\theta}}}{\cal Z}{\cal L_{\rm g}} \left( \{\omega_{AZ} \}|\bf{\theta}\right )~.
\end{equation}
 Here, ${\cal Z}$ is the evidence, $\omega_{AZ}=A Y_{AZ} / A_T$  are the mass fractions of the different clusters with mass number $A$ and charge $Z$, where $Y_{AZ}$ are the experimentally measured multiplicities and $A_T$ is the total number of nucleons in the sample. 
 $p_{{\bf\theta}}$ is the prior, and represents our knowledge of the model parameters $\theta$ before performing the Bayesian analysis; a flat prior will be taken here. Finally, for the likelihood, we consider a gaussian function $\cal L_{\rm g}$ \cite{Gelman2013}:
\begin{equation}\label{eq_likelihood}
    {\cal L_{\rm g}}(\{\omega_{AZ} \}|\theta) = \prod_{AZ} \frac{1}{\sqrt{2\pi \sigma^{2}_{AZ}}}e^{-\frac{1}{2}\left(\frac{\omega_{AZ} -m_{AZ}(\theta)}{\sigma_{AZ}}\right)^2}~,
\end{equation}
where $m_{AZ}(\theta)$ are the RMF model predictions for the mass fractions, and $\sigma_{AZ}$ the experimental uncertainties of the measured mass fractions $\omega_{AZ}$.

It is important to stress that the  event centrality and  surface velocity  $v_{\text{surf}}$ sorting described in Section \ref{sec:samples} is not an a-priori guarantee of the thermal character of the samples, and therefore of the adequacy of the description via the theoretical models of Section \ref{section:rmf_formalism}. 
The validity of the description can only be a-posteriori verified through the Bayesian analysis.
This will be done with three different checks:
\begin{itemize}
    \item The first obvious requirement is the quality of the data reproduction by the model with optimised parameters. This will be shown in Fig.\ref{fig:Mass_Fractions_xw} below.
    \item A more subtle test concerns the uniqueness of the extracted function $x_s(\rho,T)$.The presence of a large set of samples, all analyzed independently, allows to check whether two samples corresponding to the same $(\rho,T)$ value are described by compatible values of $x_s$. The absence of systematic deviations in Figure \ref{fig:T_vs_xs_xw} is a nice confirmation of this statement.
    \item Finally, the presence of dynamical effects affecting the statistical character of a specific cluster species can be directly probed by excluding this species from the Bayesian analysis and comparing the resulting theoretical yields with the experimental data. This will be done in Section \ref{section:deuterons} for the deuterons, that are more likely to be produced by coalescence and/or affected by final state interactions because of their low binding energy.
\end{itemize}

To perform the inference, we have used  
the \texttt{PyMultiNest} sampler \cite{Buchner:2014nha,Buchner:2021kpm}, which divides the posterior into nested slices with initial \textit{n-live} points, generating samples from each and recombining them to restore the original distribution \cite{Skilling2004}.

\section{Data Analysis}\label{section:data_analysis}

The main results of this work were reported in the recent publication Ref.\cite{Custodio_prl_2025}, where however many details were omitted because of length constraints.
In the next two subsections, Sec.\ref{section:alex_quadratic_fit} and \ref{section:intermediate_step}, the preliminary analyses 
leading to the work of Ref.\cite{Custodio_prl_2025} are explained and the results validated by detailed checks, starting with the approach of Refs.\cite{Pais2020,Pais2020prl}.

\subsection{Quadratic description of baryonic density and temperature}\label{section:alex_quadratic_fit}

As mentioned above, 
the data reduction provides 52 independent data sets, that a-priori could depend on 52 independent $\theta=(\rho,T,x_s)$ parameter sets. Even under the constraint that $x_s(\rho,T)$ be a unique function, this correponds to a far too large number of unknown parameters to be optimized in a Bayesian framework, and a fully agnostic analysis is not feasible. 
In previous attempts to extract the in-medium modifications of the cluster self-energies \cite{Qin2011,Pais2020,Pais2020prl}, the hypothesis was explicitly done that the reaction mechanism would correspond to an expanding cooling gas, leading to a monotonic dependence of temperature and density with the sorting parameter $v_{\text{surf}}$. This hypothesis allows a reduction of the number of parameters by assuming a parametrized behavior of $T$ and $\rho$ as \cite{Alex_Thesis_2024}:
\begin{eqnarray}\label{eq:rho_alex_quadratic}
    &&\rho(v_{\text{surf}}) = a_1  v_{\text{surf}}^2 + a_2 v_{\text{surf}} + a_3 \label{eq:rho_T_alex_quadratic}\\[5pt]
&&T(v_{\text{surf}}) = b_1 v_{\text{surf}}^2 + b_2  v_{\text{surf}} + b_3 ~.\label{eq:T_alex_quadratic}
\end{eqnarray}

As for the scalar coupling ratio $x_s$, following Refs.\cite{Pais2020,Pais2020prl}, we can at first consider a common constant value for all statistical ensembles ($v_{\text{surf}}$). Within these simplifying hypotheses, 
the number of unknown parameters is reduced to 7: $\theta = \{a_1 , a_2 , a_3 , b_1 , b_2 , b_3 , x_s\}$. These parameters can be inferred with the Bayesian analysis described in Section \ref{section:bayesian_framework}, with the likelihood {$\cal L_{\rm g}$} now taking the form:
\begin{equation}\label{eq:likelihood_alex}
    {\cal L_{\rm g}}(\{\omega_{AZ} \}|\theta) = \prod_{i,j} \frac{1}{\sqrt{2\pi \sigma_{i,j}^{2}}}e^{-\frac{1}{2}\left(\frac{\{\omega_{AZ} \}_{i,j}-m_{i,j}(\theta)}{\sigma_{i,j}}\right)^2}~,
\end{equation}
where $i=1,\dots,13$ are the different statistical ensembles ($v_{\text{surf}}$) and $j=$$^1$H,$^2$H,$^3$H,$^3$He,$^4$He the particles whose mass fractions are experimentally observed. 
The inference is repeated for all four $^{136,124}$Xe$+^{124,112}$Sn entrance channels independently, in order to provide a first test of the equilibrium hypothesis.

\begin{figure}[tp]
	\centering
		\includegraphics[width=\linewidth]{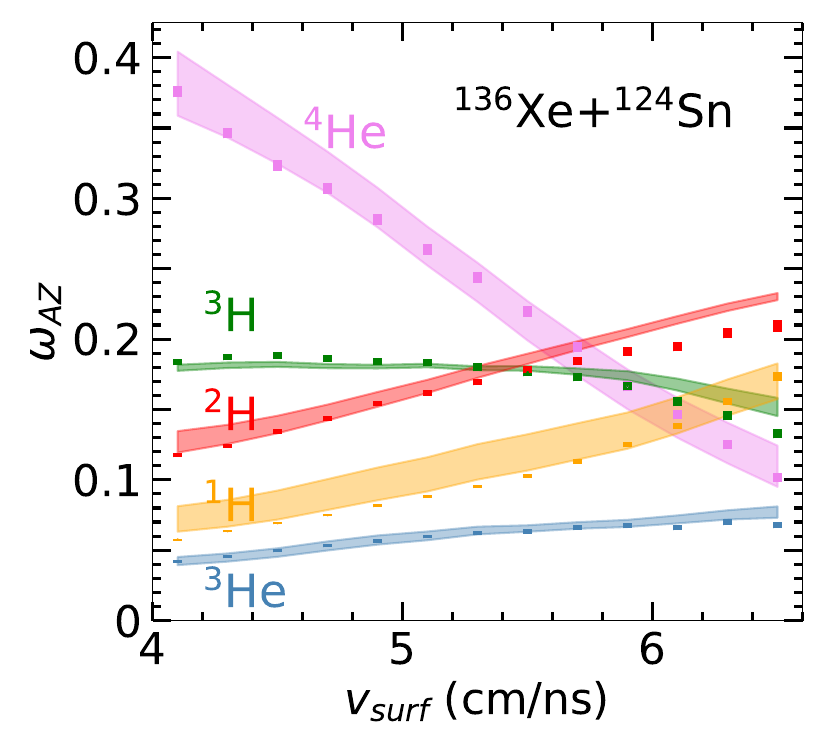}
  \caption{ \justifying Experimental (dots) and theoretical (bands) 
  mass fractions for the colliding system $^{136}$Xe$+^{124}$Sn, considering the parametric form Eqs.(\ref{eq:rho_alex_quadratic},\ref{eq:T_alex_quadratic}) for the density and temperature,
   for the FSU RMF model, according to Ref.\cite{Alex_Thesis_2024}. Both experimental dots and theoretical bands are represented with 1-$\sigma$ uncertainty. }	\label{fig:mass_fractions_alex_thesis_quadratic_fit}
\end{figure}

Figure \ref{fig:mass_fractions_alex_thesis_quadratic_fit} shows the corresponding experimental and theoretical mass fractions as a function of $v_{\text{surf}}$  for the $^{136}$Xe$+^{124}$Sn entrance channel. Even if the data reproduction is not perfect, particularly concerning the deuteron yield, a clear improvement is observed when comparing to previous works, see  
Figure 3 of the Supplemental Material of Ref.\cite{Custodio_prl_2025}, where the densities were modelled using a modified ideal gas assumption. 
A comparable level of agreement with the experimental data is observed for the other entrance channels (not shown). 
Moreover, the $x_s$ posteriors obtained for each entrance channel overlap within 1-$\sigma$ confidence interval, which is a nice confirmation of equilibrium:  $0.893^{+0.018}_{-0.016}$
($^{124}\text{Xe}+^{112}\text{Sn}$); $0.905^{+0.011}_{-0.017}$ ($^{124}\text{Xe}+^{124}\text{Sn}$); $0.899^{+0.010}_{-0.011}$ ($^{136}\text{Xe}+^{112}\text{Sn}$); $0.912^{+0.002}_{-0.007}$ ($^{136}\text{Xe}+^{124}\text{Sn}$).

\begin{figure}
	\centering
		\includegraphics[width=\linewidth]{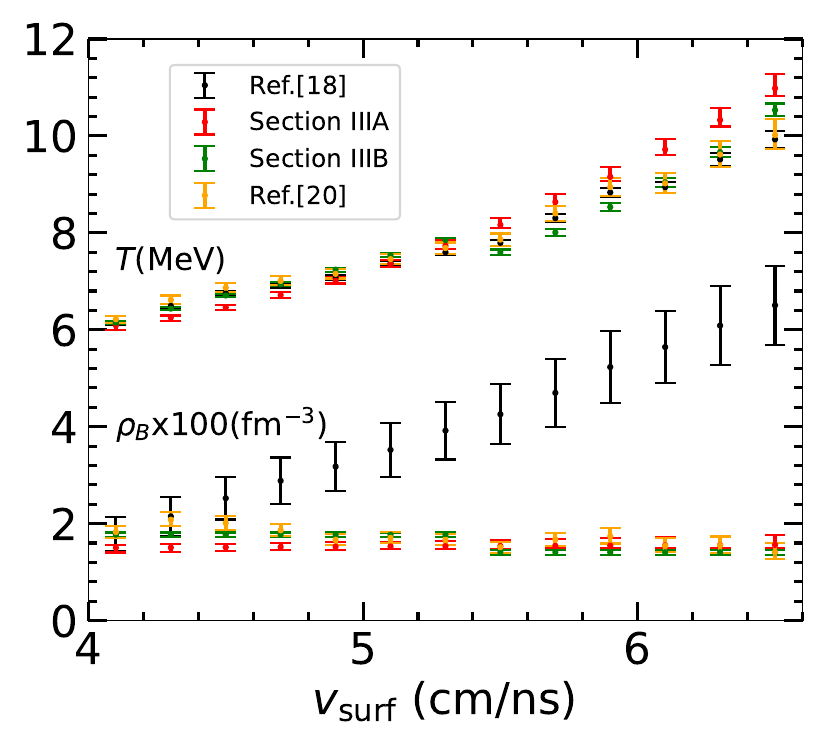}
	\caption{\justifying Comparison between extracted values of temperature and baryonic density, for the $^{136}$Xe$+^{124}$Sn entrance channel, performed considering: the modified ideal gas assumption from Refs.\cite{Pais2020,Pais2020prl} (black); density and temperature as quadratic functions of $v_{\text{surf}}$ as in Section \ref{section:alex_quadratic_fit} (red); the result of dividing the experimental points in two groups of low and high $v_{\text{surf}}$ performed in Section \ref{section:intermediate_step} (green); and the extracted density and temperature values obtained in Ref.\cite{Custodio_prl_2025} (orange). All four scenarios considered an FSU RMF model and the error bars represent 1-$\sigma$ uncertainty.
  }	
	\label{fig:rho_T_alex_thesis_quadratic_fit}
\end{figure}

However, the 
baryonic density posteriors (red symbols in Figure \ref{fig:rho_T_alex_thesis_quadratic_fit}) show a stark difference when compared to the values reported in Refs.\cite{Pais2020,Pais2020prl} (black symbols in Figure \ref{fig:rho_T_alex_thesis_quadratic_fit}). In fact, the extracted densities are compatible with a constant value of $\rho$ as a function of $v_{\text{surf}}$, whereas previous studies \cite{Pais2020,Pais2020prl} were supposing
an increasing behavior of $\rho$ as a function of $v_{\text{surf}}$, 
in line with the hypothesis of 
an expanding cooling gas. On the other hand, the temperature posteriors shown in Fig.\ref{fig:rho_T_alex_thesis_quadratic_fit}, even though not identical to those of Refs.\cite{Pais2020,Pais2020prl}, show a similar increasing trend as a function of $v_{\text{surf}}$. 

These results suggest that a better description of cluster abundances is achieved by considering that the density and temperature of each statistical ensemble are unknown and should be 
estimated in the statistical analysis instead of being theoretically modelled
as in Refs.\cite{Pais2020,Pais2020prl}. However, such an approach seems to reveal a stark contrast with the density description of previous studies, which must be investigated further.  As such, in the following subsections, we investigate whether an improvement in  data reproduction is also achieved within a more agnostic treatment beyond the hypothesis of a quadratic density and temperature, as considered in Eqs.(\ref{eq:rho_alex_quadratic},\ref{eq:T_alex_quadratic}).

\subsection{
Bayesian inference imposing a constant baryonic density
}\label{section:intermediate_step}

\begin{figure}[tp]
	\centering
		\includegraphics[width=\linewidth]{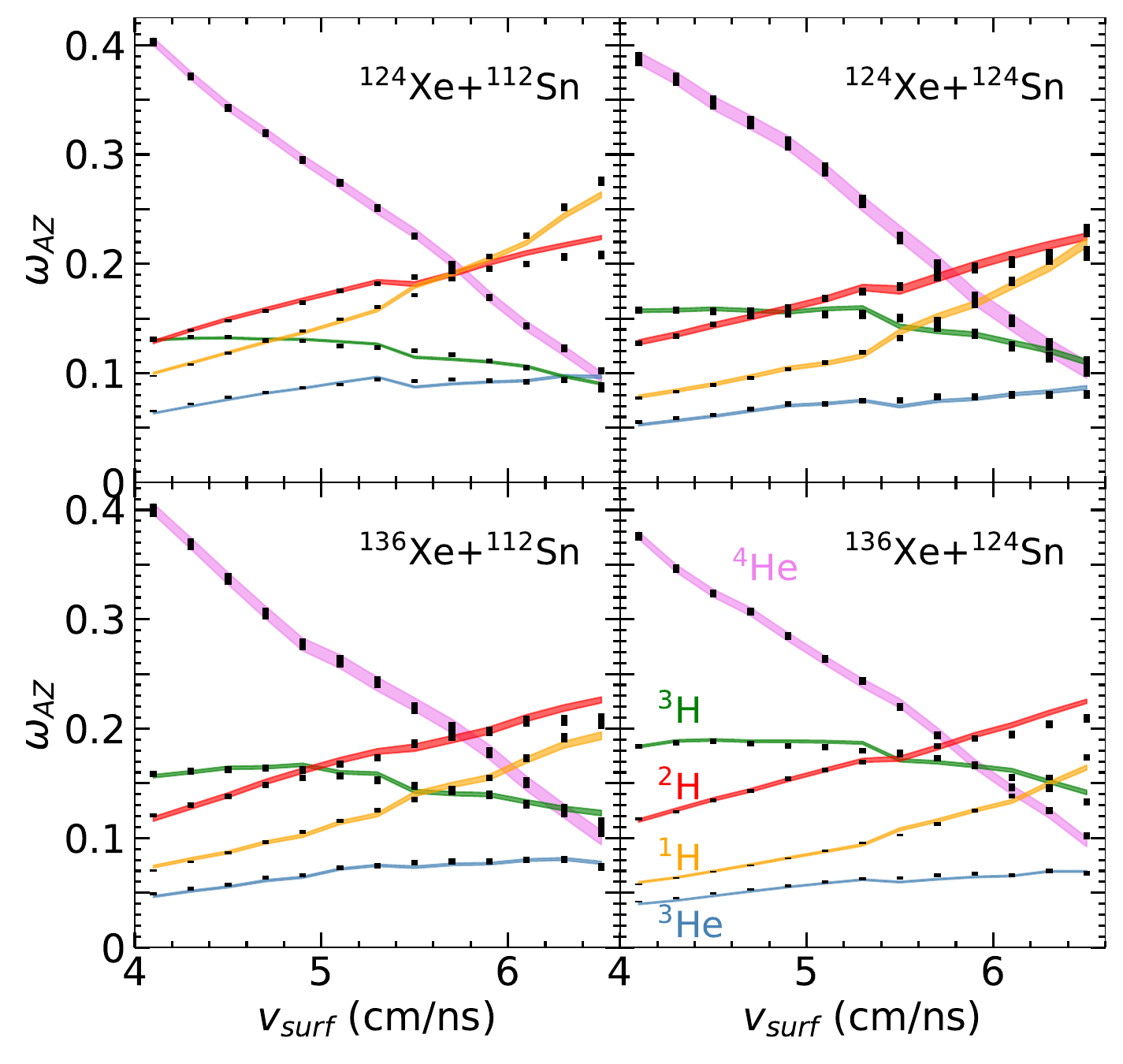}
  \caption{ \justifying Experimental (black symbols) and theoretical (bands) nuclear species mass fractions for the different colliding systems $^{136,124}$Xe$+^{124,112}$Sn and with the optimised ${\theta_1}=\{\rho_1,x_{s1},T_{4.1},T_{4.3},T_{4.5},T_{4.7},T_{4.9},T_{5.1},T_{5.3}\}$ 
   and
    ${\theta_2}=\{\rho_2,x_{s2},T_{5.5},T_{5.7},T_{5.9},T_{6.1},T_{6.3},T_{6.5}\}$ 
     parameter distributions displayed in Figs.\ref{fig:T_vs_rho_intermediate},\ref{fig:fig_xs_intermediate}, for the FSU RMF model. Both experimental black symbols and theoretical bands are represented with 2-$\sigma$ uncertainty.}	\label{fig:Mass_Fractions_intermediate}
\end{figure}

\begin{figure}[tp]
	\centering
		\includegraphics[width=\linewidth]{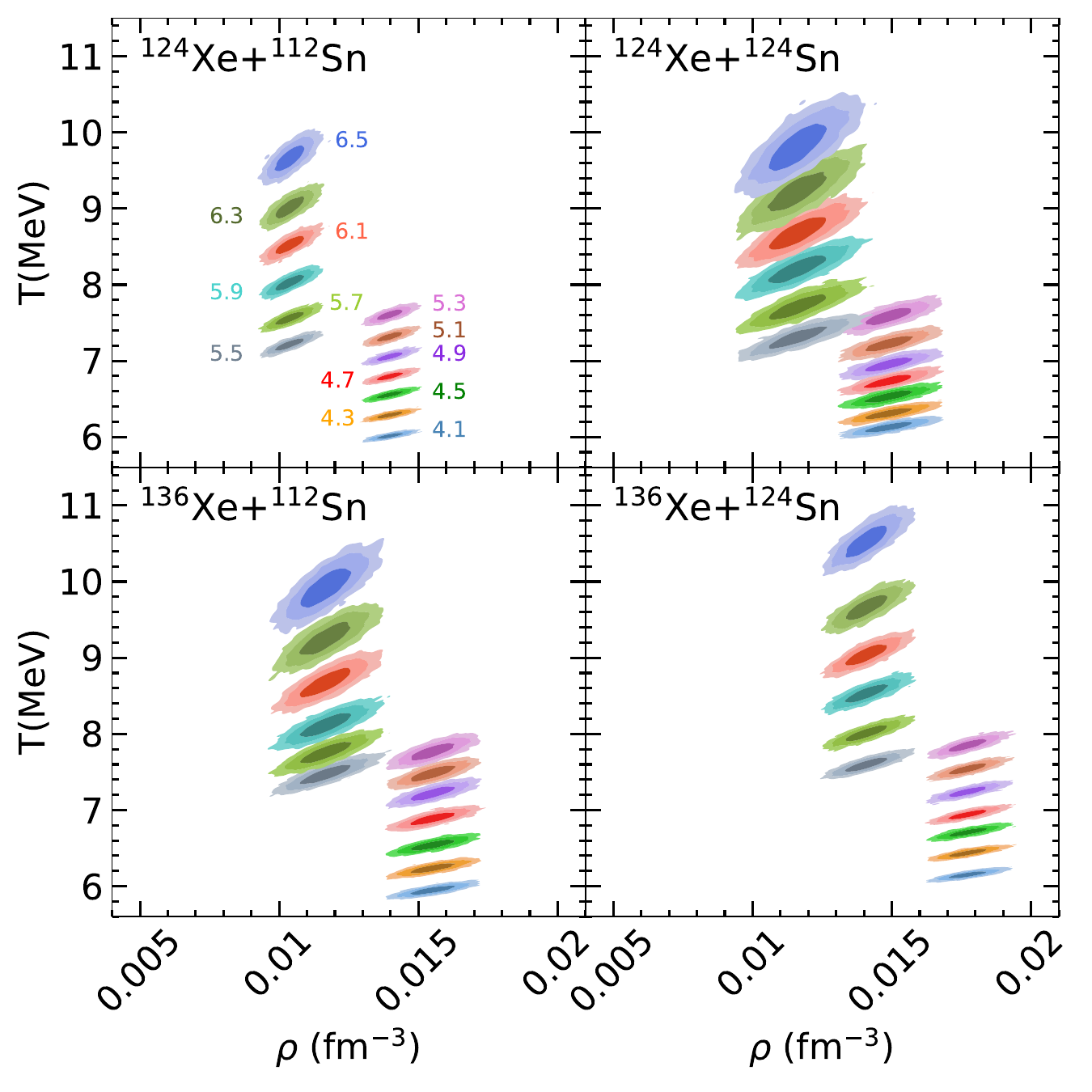}
  \caption{ \justifying Bayesian estimation of the density and temperatures belonging to the parameter sets 
  ${\theta_1}=\{\rho_1,x_{s1},T_{4.1},T_{4.3},T_{4.5},T_{4.7},T_{4.9},T_{5.1},T_{5.3}\}$ 
   and
    ${\theta_2}=\{\rho_2,x_{s2},T_{5.5},T_{5.7},T_{5.9},T_{6.1},T_{6.3},T_{6.5}\}$,
    for each entrance channel. Note that each panel includes the results of the two different bayesian inferences on ${\theta_1}$ and ${\theta_2}$, since they consider different $v_{\text{surf}}$ bins. 
  Each colour represents a different $v_{\text{surf}}$(cm/ns) bin and is identified on the upper left panel. 
  For each band, the 1,2-$\sigma$ uncertainty regions are shown for the FSU RMF model.}	\label{fig:T_vs_rho_intermediate}
\end{figure}

\begin{figure}
	\includegraphics[width=0.95
 \linewidth]{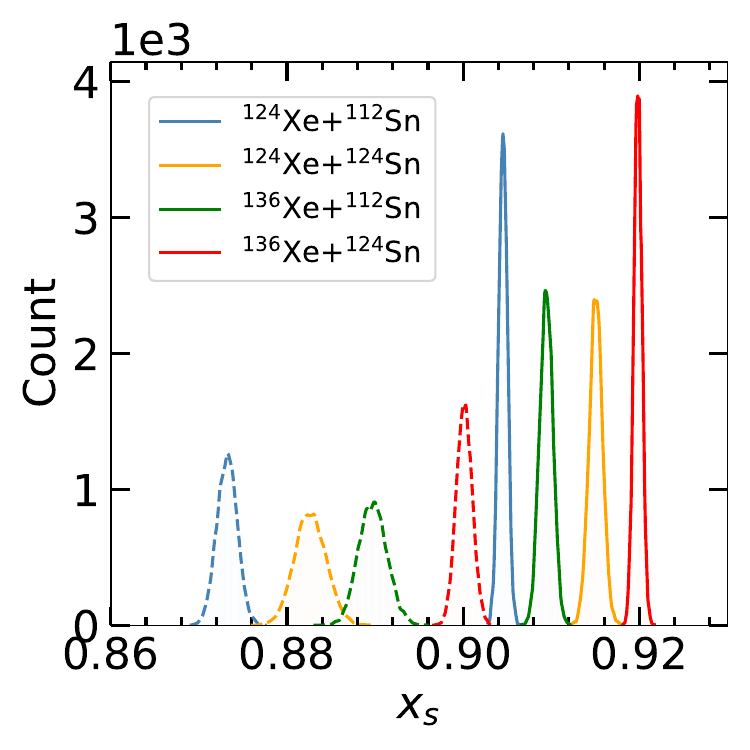} 
	\caption{\justifying 
    Posterior distributions histogram of the effective cluster coupling parameter $x_s$ corresponding to
    ${\theta_1}=\{\rho_1,x_{s1},T_{4.1},T_{4.3},T_{4.5},T_{4.7},T_{4.9},T_{5.1},T_{5.3}\}$ (solid)
   and
    ${\theta_2}=\{\rho_2,x_{s2},T_{5.5},T_{5.7},T_{5.9},T_{6.1},T_{6.3},T_{6.5}\}$ (dashed)
    parameter sets, for the FSU RMF model.
    Each colour corresponds to a different entrance channel.
    }
 \label{fig:fig_xs_intermediate}
\end{figure}

Even though in Section \ref{section:alex_quadratic_fit} there has been an improvement in the description of experimental particle abundances compared to Refs.\cite{Pais2020,Pais2020prl}, there are still some visible discrepancies. In particular:
$^1$H is slightly overestimated at low $v_{\text{surf}}$; $^2$H are overestimated at high $v_{\text{surf}}$; and $^3$H are overestimated (underestimated) for high (low) $v_{\text{surf}}$. The
parametric dependence assumed for $\rho$ and $T$ in Eqs.(\ref{eq:rho_alex_quadratic},\ref{eq:T_alex_quadratic}), 
implies a perfect bijection between the thermodynamical variables with the sorting variable $v_{\text{surf}}$, which may be an oversimplification.  

However, there seems to be a strong indication from Fig.\ref{fig:rho_T_alex_thesis_quadratic_fit} (red symbols) that the baryonic density remains approximately constant throughout the whole set of experimental points analysed.  
Because of that, in the following analysis, we consider that all $v_{\text{surf}}$ points correspond to the same baryonic density $\rho$ and ratio $x_s$ but different values for the temperature $T$, that are not necessarily correlated to $v_{\text{surf}}$. 
Due to computational limitations of PyMultinest, the 13 $v_{\text{surf}}$ bins were divided into two groups where each group is associated with a single value of the baryonic density $\rho$ and $x_s$ ratio: the first group consisting of the lowest $v_{\text{surf}}$ values, and the second with the highest $v_{\text{surf}}$ values.
Thus, for each entrance channel, two independent Bayesian analyses are performed in order to 
obtain posterior distributions of the following sets of parameters:
\begin{eqnarray}
    {\theta_1}&=&\{\rho_1,x_{s1},T_{4.1},T_{4.3},T_{4.5},T_{4.7},T_{4.9},T_{5.1},T_{5.3}\} ~~~~~~
    \\ [5pt]
    {\theta_2}&=&\{\rho_2,x_{s2},T_{5.5},T_{5.7},T_{5.9},T_{6.1},T_{6.3},T_{6.5}\} ~,~~~~~ 
\end{eqnarray}
where the temperature indices represent $v_{\text{surf}}$ values.

The likelihood is given by Eq.(\ref{eq:likelihood_alex}), 
where $i=1,2,3,4,5,6,7$ and $i=8,9,10,11,12,13$ for the $\theta_1$ and $\theta_2$ $v_{\text{surf}}$ values sets, respectively, and $j=$$^1$H,$^2$H,$^3$H,$^3$He,$^4$He.

Fig.\ref{fig:Mass_Fractions_intermediate} combines the results of these two analyses. The experimentally measured mass fractions (black symbols) are compared with the  corresponding marginalised posteriors that were obtained integrating Eq.(\ref{eq:Bayes}) 
over the $\theta_1$ and $\theta_2$ distributions. In general, a good agreement is observed between theory and experiment, except for high $v_{\text{surf}}$ values, where some deviations are visible for $^1$H, $^2$H, and $^3$He. Nevertheless, an improvement in the reproduction of $^1$H, $^2$H and $^3$H is observed when compared to Section \ref{section:alex_quadratic_fit}.

The extracted distributions of baryonic density $\rho$ and temperature $T$ are shown in Fig.\ref{fig:T_vs_rho_intermediate} for both analyses. 
 A clear distinction between the two inferences seems to indicate a density 
dependence for the selected $v_{\text{surf}}$ samples. This dependence
is also observed in Fig.\ref{fig:rho_T_alex_thesis_quadratic_fit} where the green data points correspond to the same analyses. 
It is however important to stress that the samples span a large temperature interval, with the first one exploring systematically lower $T$ values than the second one. If the coupling constant is temperature dependent, our assumption of a unique $x_s$ value over the samples of each inference could bias the density extraction. All in all,
although a higher density is obtained for lower $v_{\text{surf}}$, the extracted density values are still consistent with those obtained for a constant density in Section \ref{section:alex_quadratic_fit} within the error bars of that analysis shown in  Fig.\ref{fig:rho_T_alex_thesis_quadratic_fit} by the red data points. These results contrast  with the expanding cooling gas picture  of previous works \cite{Pais2020,Pais2020prl}, shown with the black data points in the same figure. On the other hand, Fig.\ref{fig:rho_T_alex_thesis_quadratic_fit} suggests that the temperature profile increases with $v_{\text{surf}}$, as observed in all scenarios.

The most compelling consistency test of the analysis is presented in Fig.\ref{fig:fig_xs_intermediate}, where the posterior
distributions for the $x_s$ ratio within the two sets $\theta_1$ and $\theta_2$, are plotted  for the different $\text{Xe}+\text{Sn}$ systems. 

If everything is under control, the extracted coupling $x_s$ should be unique, which is clearly not the case in Fig.\ref{fig:fig_xs_intermediate}.
In fact, the $x_s$ ratio is systematically larger in the lower $v_{\text{surf}}$ samples (solid lines in Fig.\ref{fig:fig_xs_intermediate}). 

Since the temperature increases significantly with $v_{\text{surf}}$, this might be an indication that $x_s$ has some 
temperature dependence.
However, a small but monotonic and systematic increase of $x_s$ with decreasing $v_{\text{surf}}$ is also observed within each entrance channel (different colours in Fig.\ref{fig:fig_xs_intermediate}).
As it can be appreciated from Fig.\ref{fig:T_vs_rho_intermediate}, within a same $\theta$ group, the different entrance channels span slightly different density and, to a minor extent, temperature ranges. The residual difference between the $x_s$ values of the different systems within a same $v_{\text{surf}}$ interval might  then signal some residual density dependence. Since the difference between the $x_s$ values extracted from the samples differing in density (e.g. dashed blue and dashed red in Fig.\ref{fig:fig_xs_intermediate}) is very close to the difference between the $x_s$ values extracted from the samples differing in temperature (e.g. dashed blue and full blue in Fig.\ref{fig:fig_xs_intermediate}), it is very difficult to discriminate between the two hypotheses. However, the density extraction in our inferences appears less robust than the temperature extraction, and posterior density values fluctuate within the $(0.01,0.02)$ fm$^{-3}$ interval depending on the different technical details of the sampling. For this reason, we do not think we are able to robustly discriminate density values within this interval, and the hypothesis of a dominant temperature dependence of the coupling appears better justified by the analysis results. 

Finally, it is also interesting to compare the results obtained within the two entrance channels associated to the same total baryonic mass $^{124}$Xe$+^{124}$Sn and $^{136}$Xe$+^{112}$Sn (upper right and lower left panel in Fig.\ref{fig:T_vs_rho_intermediate}, and yellow and green, respectively, in Fig.\ref{fig:fig_xs_intermediate}). We observe that the low  and high $v_{\text{surf}}$ samples $\theta_1$ and $\theta_2$ span clearly distinguished distinct density and temperature values, which however are sensibly the same between the two entrance channels. We would therefore expect very close $x_s$ values for the two entrance channels, i.e., a huge superposition between the yellow and green $x_s$ posteriors. 
This is not observed in Fig.\ref{fig:fig_xs_intermediate}.
A lower $x_s$ value is obtained in the isospin symmetric system with respect to the isospin asymmetric one in the high temperature, low density sample (yellow dashed versus green dashed), while the opposite is true in the low temperature, high density sample (yellow solid versus green solid). 
The different ordering of the yellow and green distributions between the low   and high $v_{\text{surf}}$  samples indicates that the widths of the $x_s$ distributions should be interpreted with care, and the uncertainty in the $x_s$ extraction due to uncontrolled parameters and out-of-equilibrium effects is of the order of the difference between these two posteriors, $\Delta x_s\approx 0.01$.
This is  consistent with the 2-$\sigma$ confidence interval obtained for  the $\sigma$-cluster and  the $\omega$-cluster shown in Fig.  \ref{fig:fit_xs_xw_FSU_DD2}.

With these caveats in mind, 
an independent Bayesian analysis was performed in Ref.\cite{Custodio_prl_2025} to calibrate ($T$,$\rho$,$x_s$) for each $v_{\text{surf}}$ bin and entrance channel. In that work, a temperature dependence of $x_s$ was  observed, consisting in a decrease of $x_s$ for increasing temperature values. On the other hand, the extracted baryonic densities were compatible with a single value of $\rho \sim$0.015 fm$^{-3}$. 
Our estimates $\Delta \rho \approx 0.01$ fm$^{-3}$, $\Delta x_s\approx 0.01$ are fully consistent with the results of Ref.\cite{Custodio_prl_2025} (see the posterior dispersion of these variables in Fig.\ref{fig:T_vs_rho_xw} and \ref{fig:T_vs_xs_xw} below).
In the following Section \ref{section:xw_calibration}, we will repeat the Bayesian inference done in \cite{Custodio_prl_2025},  taking into account the degeneracy between the cluster coupling to the $\sigma$ and $\omega$-meson: instead of fixing $x_\omega$,  we fix $x_s$ and allow the coupling ratio $x_\omega$ to vary with temperature and density. One reason for inferring $x_\omega$ instead of $x_s$ is that extrapolating $x_s$ for temperatures above 10 MeV may result in negative values.  As temperature increases, we expect less binding and therefore more repulsion. Therefore, we may consider the $\omega$-meson to describe this effect.

\subsection{
Evaluation of the degeneracy between $x_s$ and $x_{\omega}$
}\label{section:xw_calibration}

\begin{figure}[tp]
	\centering		\includegraphics[width=\linewidth]{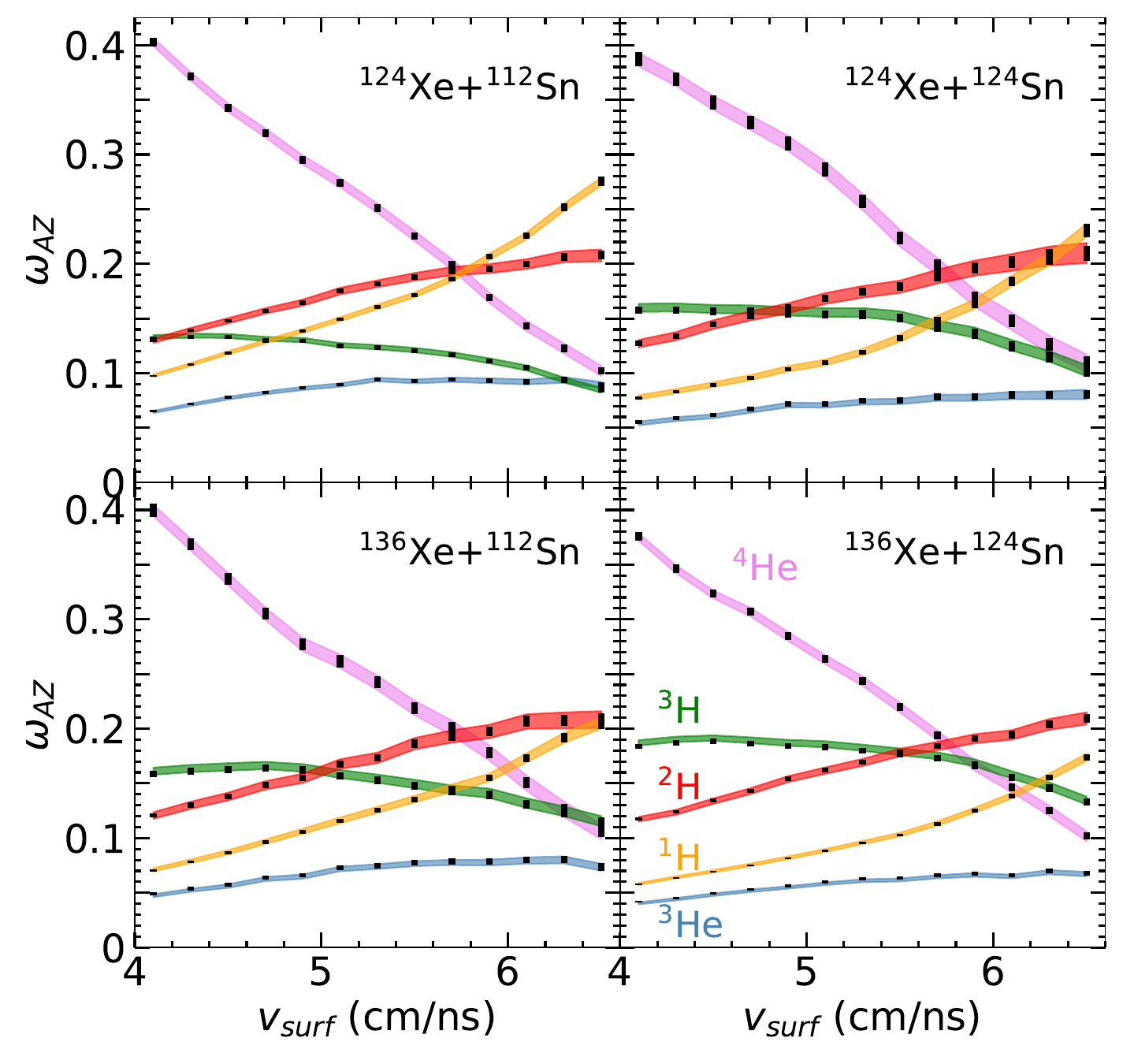}
  \caption{ \justifying Experimental (black symbols) and theoretical (bands) nuclear species mass fractions for the different colliding systems $^{136,124}$Xe$+^{124,112}$Sn and  with the optimised $(\rho, T, x_{\omega})$ parameter distributions displayed in Figs.\ref{fig:T_vs_xs_xw},\ref{fig:T_vs_rho_xw}, for FSU RMF model. Both experimental black symbols and theoretical bands are represented with 2-$\sigma$ uncertainty. }	\label{fig:Mass_Fractions_xw}
\end{figure}

\begin{figure}[tp]
	\centering
		\includegraphics[width=1.\linewidth]{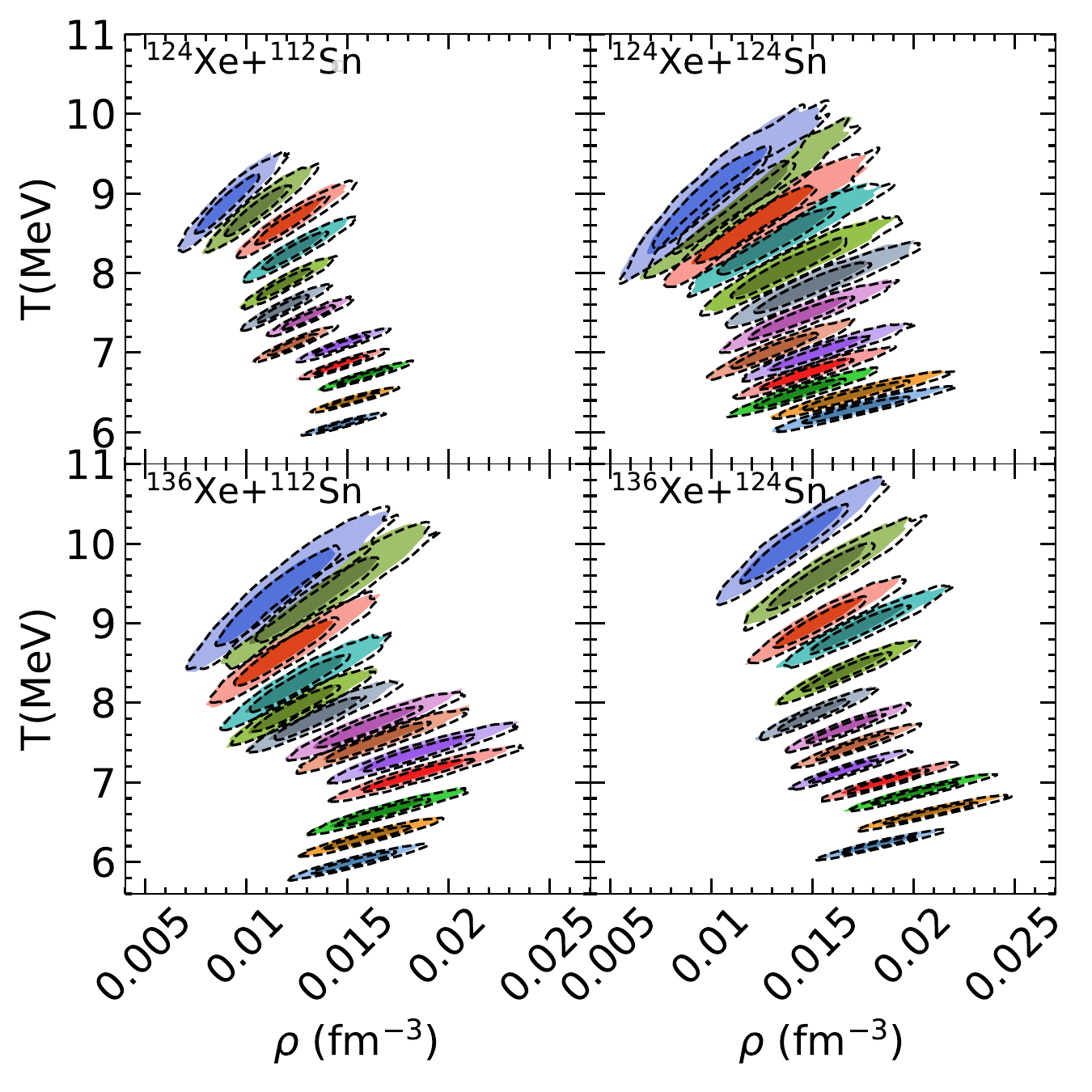}
  \caption{ \justifying Bayesian estimation of the thermodynamical parameters $\rho$ and $T$ for the cases where $x_s$ \cite{Custodio_prl_2025} (filled contours) or $x_{\omega}$ (unfilled contours delimited with grey dashed lines) were considered in the calibration. 
  Each colour represents a different $v_{\text{surf}}$(cm/ns) bin and is identified on the upper left panel. 
  For each band, the 1,2-$\sigma$ uncertainty regions are shown for FSU RMF model.} 
	\label{fig:T_vs_rho_xw}
\end{figure}

The extracted ($T$,$\rho$,$x_s$) values in Ref.\cite{Custodio_prl_2025} allowed for an excellent reproduction of the experimental data, showing a clear improvement compared to previous studies \cite{Pais2020,Pais2020prl}. As expected,  the results of Ref.\cite{Custodio_prl_2025} also show a much better description of the experimental mass fractions than those presented in Sections \ref{section:alex_quadratic_fit} and \ref{section:intermediate_step} of the present work. This translated into a similar temperature profile as observed in Fig.\ref{fig:rho_T_alex_thesis_quadratic_fit} (orange), while also being consistent with a single 
 density value, even though a slight density decrease with  $v_{\text{surf}}$, as the  one  found in Sec. \ref{section:intermediate_step}, is also observed in \cite{Custodio_prl_2025}. Concerning the coupling $x_s$, a temperature dependence was established in \cite{Custodio_prl_2025}, while a density dependence was inconclusive due to the very slight variation of the extracted densities throughout the considered experimental points.

The improvement observed in \cite{Custodio_prl_2025} was possible because the density was directly extracted from the data, without assuming any modified ideal gas hypothesis, 
as it is routinely done in parameter estimation studies with Bayesian inference.

In principle, cluster quenching in the medium could be modelled both by increasing the cluster effective mass  with respect to the standard mean field value (that is, introducing a value $x_s< 1$), or by increasing the vector repulsion for nucleons bound into a cluster (that is, introducing a value $x_\omega> 1$). 
So far, in both the present paper and in Ref. \cite{Custodio_prl_2025}, it was chosen to fix the cluster coupling to the $\omega$ meson to the value $x_{\omega}=1$ and estimate 
the cluster coupling to the $\sigma$ meson by introducing an $x_s< 1$ ratio. 
Then, the question arises about the uniqueness of this ansatz, and whether our posteriors for the densities, temperatures, and mass fractions would depend on this choice.

Thus, in the following, we repeat the analysis performed in Ref.\cite{Custodio_prl_2025},  
setting $x_s=1$ and 
treating the $x_{\omega}$ ratio as a free unknown parameter. In Fig.\ref{fig:Mass_Fractions_xw}, we 
present the particle  mass fractions as a function of $v_{\text{surf}}$, obtained 
varying $x_\omega$.
In the range of 
the data, both scenarios produce the same outcome and it is impossible to say which one is the correct approach, since both of them are able to perfectly reproduce the experimental mass fractions, as displayed in Fig.\ref{fig:Mass_Fractions_xw} for $x_{\omega}$ and in  Fig. 1 of Ref.\cite{Custodio_prl_2025} for $x_s$.

Interestingly, the degeneracy between both inferences
does not introduce any ambiguity in the determination of the thermodynamical parameters,
as shown in Fig.\ref{fig:T_vs_rho_xw}, where results corresponding to the cases where $x_s$ (filled contours) and $x_{\omega}$ (unfilled contours delimited with grey dashed lines) are displayed. 
The extracted thermodynamic properties thus appear as robustly inferred from the analysis.

\begin{figure}
	\centering\includegraphics[width=0.95
 \linewidth]{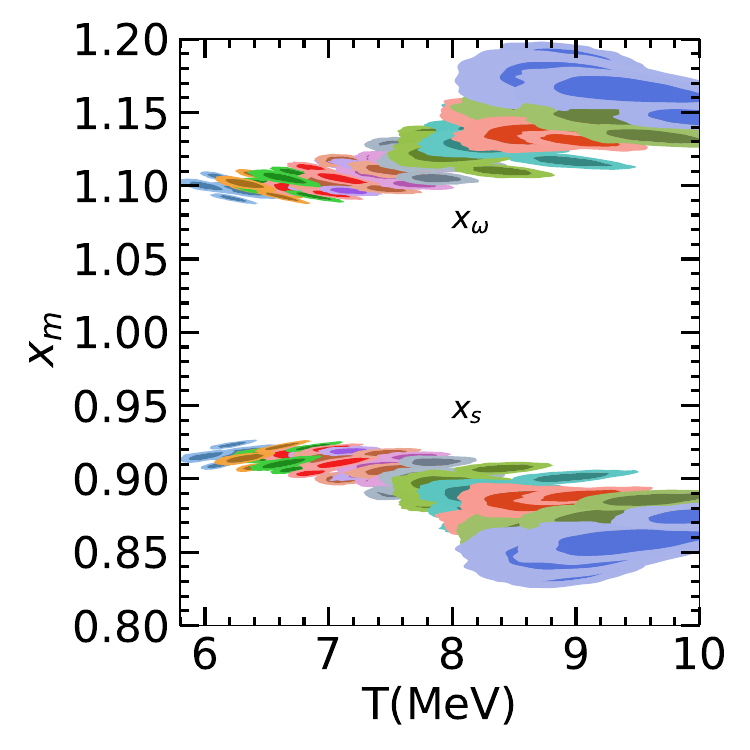} 
	\caption{\justifying  Posterior distributions of the effective cluster coupling ratios $x_s$ (bottom) and $x_{\omega}$ (top) as a function of the temperature $T$ 
    for the FSU RMF model.
    Each colour corresponds to a specific $v_{\text{surf}}$ bin as in Fig. \ref{fig:T_vs_rho_intermediate} and the four distributions per colour correspond to the four independent inferences performed on the different colliding systems. For each $v_{\text{surf}}$,  1,2-$\sigma$ uncertainty regions are shown.
    }
 \label{fig:T_vs_xs_xw}
\end{figure}

In Fig. \ref{fig:T_vs_xs_xw}, the  extracted values of $x_s$ (bottom) or of $x_{\omega}$ (top)  are both plotted as a function of the temperature  for all entrance channels. 
In both cases, as discussed in detail in the previous section, the results from the different entrance channels are compatible within the precision of the inference, which is a nice confirmation of the validity of the statistical equilibrium framework of the analysis.
The degeneracy between the scalar and vector couplings in the RMF framework is apparent from the figure. 
The decrease of $x_s$ with increasing $T$ of the inference presented in the previous section is here replaced by an increase of $x_{\omega}$. In other terms, the binding energy shift predicted by microscopic calculations\cite{Ropke:2011tr,Ropke2015,Ropke_2020,Ren_2024}
can be seen in RMF terms as a
weakening of the $\sigma$-cluster coupling with temperature 
or alternatively, as an increase of repulsion given by an increasing $\omega$-cluster coupling.

\begin{figure}
	\includegraphics[width=0.95
 \linewidth]{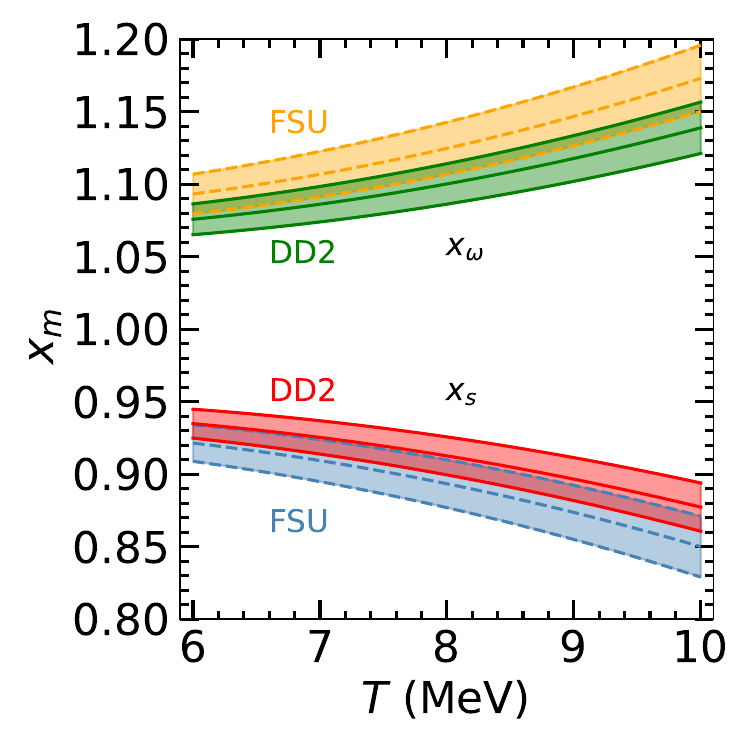} 
	\caption{\justifying Quadratic fits to model the temperature dependence of the $\sigma$-cluster coupling ratio $x_s=aT^2 + bT + c$ and $\omega$-cluster coupling ratio $x_{\omega}=aT^2 + bT + c$, taking into account all the ($x_s$,$T$) and ($x_{\omega}$,$T$) Bayesian calibrated values for the four entrances channels, for the DD2 (filled bands with solid line contours) and FSU (unfilled bands with  dashed contours) RMF models. For each $x_m$ and RMF model, both the median and 2-$\sigma$ confidence interval are shown.
    }
 \label{fig:fit_xs_xw_FSU_DD2}
\end{figure}

\begin{figure}
	\centering\includegraphics[width=0.95
 \linewidth]{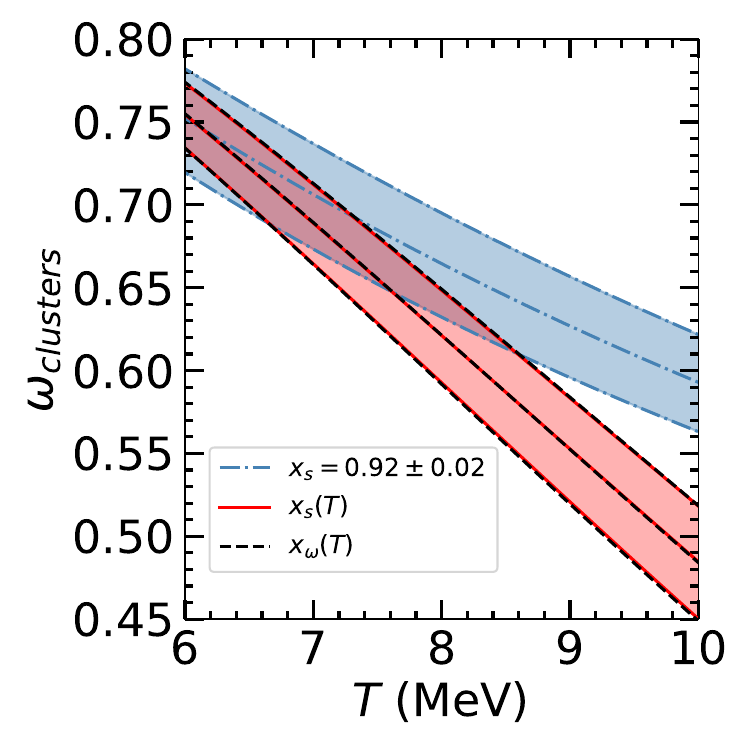} 
	\caption{\justifying  
    Total mass fraction predictions as a function of the temperature obtained
with the FSU RMF model considering the temperature dependence of the $\sigma$-cluster coupling ratio $x_s(T)$ as in Ref.\cite{Custodio_prl_2025} (filled red band with  red solid line contours) and the temperature dependence of the $\omega$-cluster coupling ratio $x_{\omega}(T)$ (unfilled band with
black dashed contours). The proton fraction $y_q=0.45$ and baryonic density $\rho=0.015$ fm$^{-3}$ were fixed. These results are compared to the mass
fractions determined with FSU and a temperature-independent value 
$x_s=0.92\pm0.02$ from Refs.\cite{Pais2020,Pais2020prl} (blue band
with dash-dotted delimiters). Both the median and 2-$\sigma$ confidence interval are shown for each case.
    } \label{Mass_Fractions_FSU_PRL_2020_xs_xw}
\end{figure}

As in Ref.\cite{Custodio_prl_2025}, where a quadratic fit was performed to model the evolution of $x_s$ with $T$, the same can be done to describe the temperature dependence of $x_{\omega}$. The parameters of the fitting, $x_{\omega} = aT^2 + bT + c$, are given in Table \ref{table_parameters_xw_fit} for both FSU and DD2 RMF models. In Fig.\ref{fig:fit_xs_xw_FSU_DD2}, the quadratic fits of $x_{\omega}$ obtained here and the analogous $x_s$ fits of Ref.\cite{Custodio_prl_2025} are shown for both the DD2 (filled bands with solid line contours) and FSU (unfilled bands with dashed contours) RMF models. The temperature dependence of both $x_s$ and $x_{\omega}$ is well accounted for by their corresponding quadratic fits. 

It is interesting to observe that even though both FSU and DD2 are able to produce very similar extracted values for the densities and temperatures, their corresponding cluster-meson ratios 
are not the same, with DD2 showing higher (lower) values for the $x_s$ ($x_{\omega}$).
This is not surprising because the different phenomenological RMF functionals by definition correspond to different nuclear models that are equally able to describe the data. As it has
already been documented in Ref.\cite{Custodio2020}, 
the $x_s$ and $x_{\omega}$ ratios are model dependent as are the nucleonic couplings $g_s$ and $g_\omega$, but once fixed
to reproduce experimental data,  the model dependence is removed, and the  thermodynamical variables such as the baryonic density and temperature associated to a specific distribution of particle mass fractions are robustly inferred.

To analyse how the temperature dependence of $x_s$ and $x_{\omega}$ influence the light clusters abundances and better show the model independence of the results, after calibration to the data, in Fig.\ref{Mass_Fractions_FSU_PRL_2020_xs_xw} we compare the total light cluster abundances predicted by the FSU RMF model when considering the following scenarios: $x_s(T)$ from Ref.\cite{Custodio_prl_2025} (filled red band with
red solid line contours); $x_{\omega}(T)$ from this work (unfilled band with
black dashed contours); 
and $x_s=0.92\pm 0.02$ from Refs.\cite{Pais2020,Pais2020prl} (blue band with dash-dotted
delimiters). A total proton fraction of $y_p=0.45$ was taken as an example within the range of values explored by the four entrance channels and a fixed baryonic density of $\rho=0.015$ fm$^{-3}$ was considered. As expected, the bands corresponding to a temperature dependent $x_s$ and $x_{\omega}$ produce essentially the same particle abundances. The differences between these two predictions and the one of Refs.\cite{Pais2020,Pais2020prl} appears for higher temperature values, where the weakening of the cluster couplings either through a decreasing $x_s$ or an increasing $x_{\omega}$ results in smaller abundances due to less bound clusters. 

\begin{figure*}
	\centering
		\includegraphics[width=0.8\linewidth]{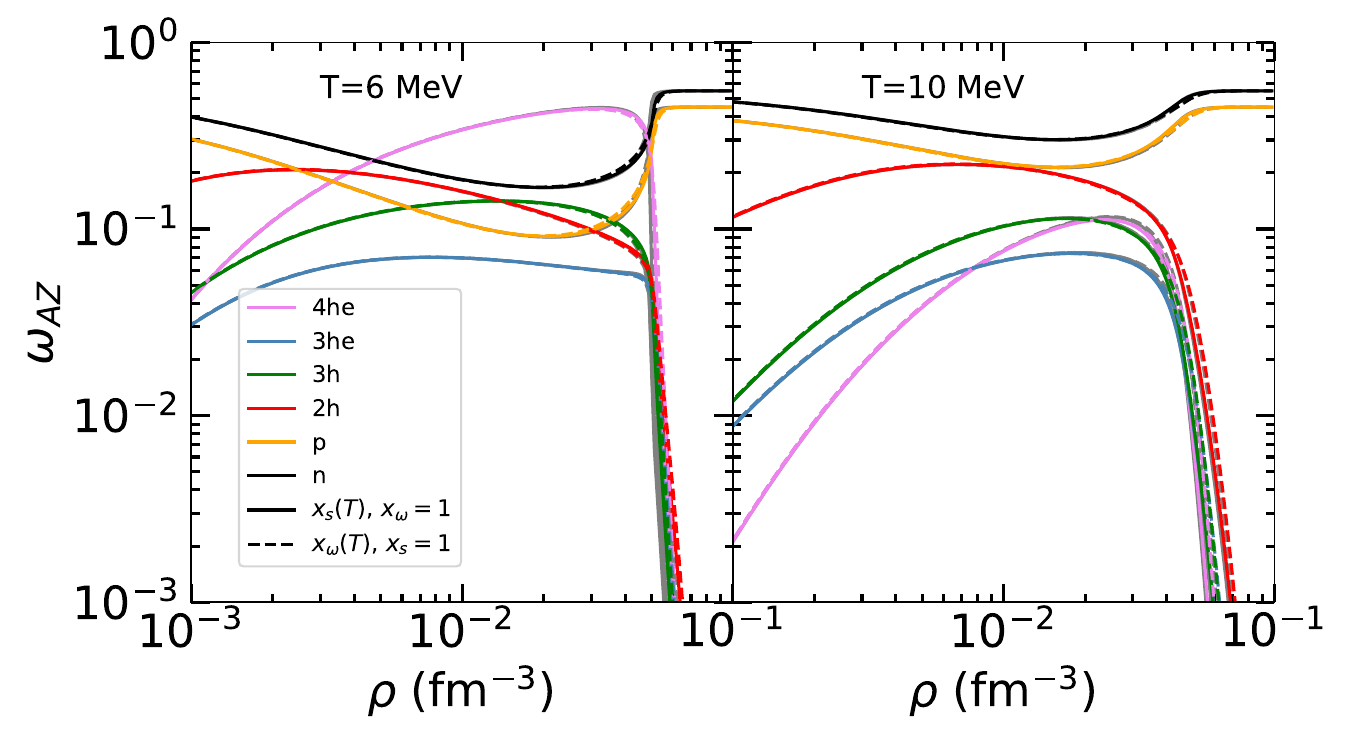}
	\caption{\justifying Comparison between particles mass fractions as a function of the baryonic density obtained considering the temperature dependence of the $\sigma$-cluster coupling ratio $x_s(T)$ as in Ref.\cite{Custodio_prl_2025} (solid) and the temperature dependence of the $\omega$-cluster coupling ratio $x_{\omega}(T)$ (dashed), for the FSU  (color lines) and DD2 (grey lines) RMF models, for fixed proton fraction $y_q=0.45$ and $T=6$ MeV (left), $T=10$ MeV  (right).
  }	
	\label{Mass_Fractions_FSU_xs_xw_median}
\end{figure*}

To further explore the two different $x_s$ and $x_{\omega}$ approaches and their influence in particle abundances, in Fig.\ref{Mass_Fractions_FSU_xs_xw_median} we compare the individual particle mass fractions as a function of baryonic density, obtained with the FSU (colored lines) and DD2 (grey lines) RMF models. Results are shown considering either the temperature dependent $x_s(T)$ of Ref.\cite{Custodio_prl_2025} (solid lines) or the $x_{\omega}(T)$ (dashed lines). For that, as an example, the proton fraction was once again fixed $y_p=0.45$ and two 
different temperature values 
were considered (marking the boundaries of temperatures extracted from the data). We notice that both approaches 
lead to very similar results for the individual particle mass fractions. Only at densities close to cluster dissolution small discrepancies start to show, being more visible at higher temperatures. Considering that this 
density range falls beyond the values explored by the data, and therefore  corresponds to an extrapolation of the fitted values, the results can be considered as satisfactory and the model dependence as weak.

\begin{table}[h]
\centering
		\caption{\justifying Parameter estimates $a, b, c$ for FSU and DD2 with 1, 2$\sigma$ uncertainties for the model $x_{\omega} = aT^2 + bT + c$, where $T$ is measured in MeV.}
		\label{table_parameters_xw_fit}
  \setlength{\tabcolsep}{5.5pt}
      \renewcommand{\arraystretch}{1.5}
\begin{tabular}{ccccc}
\hline \hline
Parameter & Unit & Median & $1\sigma$ & $2\sigma$  \\ \hline
\hspace{-0.74cm}FSU $a$ & MeV$^{-2}$ & $0.00210$ & $\pm{0.00004}$ & $\pm{0.00008}$ \\
$b$ & MeV$^{-1}$ & $-0.01370$ & $\pm{0.00057}$ & $\pm{0.00113}$ \\
$c$ & & $1.09969
$ & $\pm{0.00219}$ & $\pm{0.00433}$ \\
\\
\hspace{-0.74cm}DD2 $a$ & MeV$^{-2}$ & $0.00179$ & $\pm{0.00003}$ & $\pm{0.00006}$ \\
$b$ & MeV$^{-1}$ & $-0.01285$ & $\pm{0.00044}$ & $\pm{0.00087}$ \\
$c$ & & $1.08846$ & $\pm{0.00171}$ & $\pm{0.00336}$ \\
\hline

\end{tabular}
\end{table}

\section{Estimating out-of-equilibrium effects 
}\label{section:deuterons}

The difference  between the results obtained in this work and in Ref.\cite{Custodio_prl_2025} with respect to the results of previous studies \cite{Qin2011,indra,Pais2020prl,Pais2020}, and in particular the different estimation of the density
may come as a surprise and raise doubts as to whether these discrepancies might be caused by systematic errors due to 
out-of-equilibrium effects. The consistency between the $x_s(T)$ or $x_\omega(T)$ between the different reacting systems shown in Fig.\ref{fig:T_vs_xs_xw} above is an encouraging sign of the absence of entrance channel effects. However, dynamical effects, not captured by our statistical analysis, could systematically affect one (or more) of the detected particle species. Particularly, following Qin {\it et al.} \cite{Qin2011}, one could argue that the deuteron, due to its very low binding energy, could undergo final-state interactions that might not respect the detailed balance implied by the equilibrium hypothesis. This would lead to a deuteron multiplicity systematically lower than the equilibrium prediction, if the dominant mechanism is deuteron break-up, or to a higher one, if the dominant mechanism is coalescence.

In this section, we make the explicit hypothesis that the detected deuteron multiplicity may be systematically different from the equilibrium prediction, and we exclude the deuteron data from the thermodynamic inference. We analyse how the results obtained in \cite{Custodio_prl_2025}  would change under this hypothesis. 
We focus on the deuteron case for the aforementioned reasons, but in principle, the analysis could be repeated by excluding other particle species. Whatever the piece of data excluded from the Bayesian constraint, the corresponding posterior distribution should be seen as a theoretical prediction for that observable, and the level of adequacy between this prediction and the measured data will therefore measure the degree of reliability of the underlying model and the robustness of the inference. In this sense, the analysis presented in this section can be seen as a general example of 
how reducing the available information affects the inference of unknown parameters in a Bayesian scheme.

\begin{figure}
	\centering\includegraphics[width=
 \linewidth]{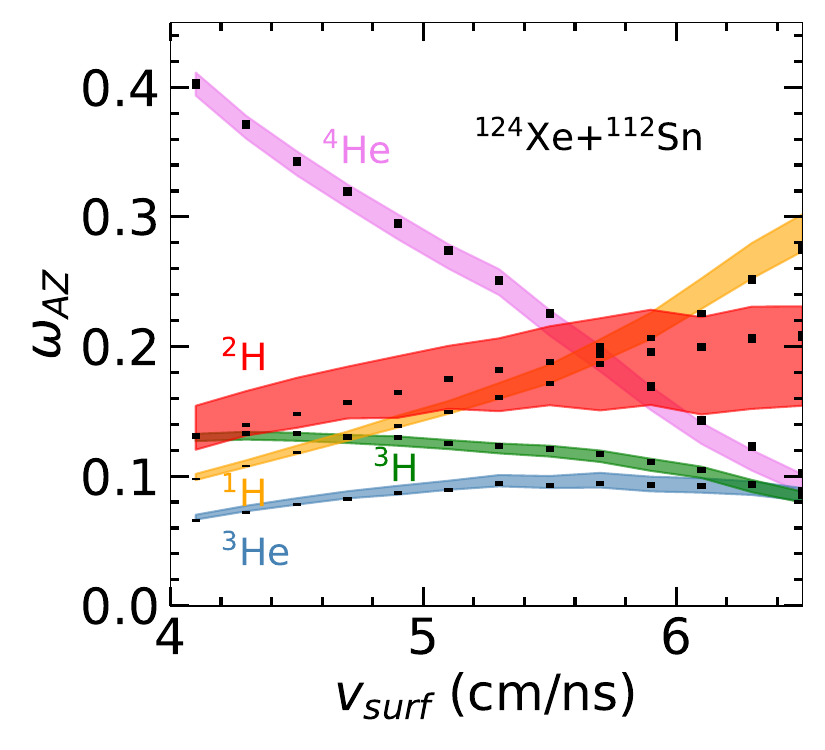} 
	\caption{\justifying Experimental (black symbols) and theoretical (bands) nuclear species mass fractions for the colliding system $^{124}$Xe$+^{112}$Sn with the optimised $(\rho, T, x_s)$ parameter distributions obtained without using the deuteron data in the Bayesian inference, for FSU RMF model. Both experimental black symbols and theoretical bands are represented with 2-$\sigma$ uncertainty. 
    }
 \label{fig:Mass_Fractions_no_2h_in_likelihood_but_comparing_to_mass_fractions_with_2h}
\end{figure}

\subsection{Methodology}

Similarly to Sec. \ref{section:data_analysis}, in principle we would have four unknown parameters in the statistical analysis, namely ($T$, $\rho$, $y_p$, $x_s$): three thermodynamic variables and the scalar coupling of the clusters which may depend on the thermodynamic variables. However, we have seen that the experimental estimate of $y_p$, obtained through the experimental estimation of neutron multiplicities from the proton spectra, leads to an excellent reproduction of the $^3\text{H}$/$^3\text{He}$ ratio in the RMF model in both Sec. \ref{section:xw_calibration} and in \cite{Custodio_prl_2025}. Therefore, in the following, we will use the same $y_p(v_{\text{surf}})$ as in Sec. \ref{section:data_analysis} and in \cite{Custodio_prl_2025}. For the same reason, there is no indication that an isospin dependence is needed in the in-medium binding energy shift. Thus, the parameters to be inferred are as in the previous section $\theta=\{T,\rho,x_s(T, \rho)\}$. It should be noted that, if the deuteron data is to be excluded from the sample, we expect that the
temperature extracted using the Albergo formula \cite{albergo} (which in all previous analyses of both INDRA \cite{indra,Pais2020prl,Pais2020,Custodio_prl_2025} and TAMU \cite{Qin2011} data was either assumed true or shown to be in good agreement with theoretical calculations) should be considered as a biased estimator, meaning that a Bayesian inference of $T$ becomes essential.

Once again, the data to be fitted in  the Bayesian analyses are the mass fractions $\omega_{\text{AZ}}$. However, if we exclude the deuteron from the analyses, the mass fractions must be renormalised accordingly. For both the experimental and the theoretical sample, we introduce the reduced mass fractions derived from the measured multiplicities {$Y_{\text{AZ}}$} for a particle with mass number $A$ and proton number $Z$ as:
\begin{equation}
    \omega_{\text{AZ}}^{\text{red}}={\frac{A\, Y_{\text{AZ}}}{\sum_j A_j\, Y_j}}~,\,j=n,^1\text{H},^3\text{H},^3\text{He},^4\text{He} ~. 
\end{equation}

Since the Bayesian analyses are now performed on the reduced mass fractions, the likelihood ${\cal L_{\rm g}}$ becomes:
\begin{equation}
    {\cal L_{\rm g}}(\{\omega_{AZ}^\text{red} \}|\theta)= \prod_{i,j} \frac{1}{\sqrt{2\pi \sigma_{i,j}^{2}}}e^{-\frac{1}{2}\left(\frac{\omega_{i,j}^{\text{red,RMF}} -\omega_{i,j}^{\text{red,exp}}}{\sigma_{i,j}}\right)^2} ~.
\end{equation}

It is important to notice that, even though the deuteron data was removed from the Bayesian inference, the deuteron channel should not be removed from the thermodynamic equilibrium calculated with RMF. As such, for the optimised values of $\theta=\{T,\rho,x_s(T, \rho)\}$, we can calculate the mass fractions $\omega_{\text{AZ}}$ including the deuteron contribution. This is precisely what we show in Fig.\ref{fig:Mass_Fractions_no_2h_in_likelihood_but_comparing_to_mass_fractions_with_2h}, where 
the comparison between the posterior distributions of the calculated mass fractions $\omega_{\text{AZ}}$ 
and the experimentally measured ones are plotted. We see that the degree of reproduction of the data is slightly worse than the one of \cite{Custodio_prl_2025}, as expected if the extra deuteron constraint bears a significant information, but still acceptable.

The posterior distribution of deuterons,  whose data were not imposed in the Bayesian  fit, should now be interpreted as a prediction of the model, rather than as a measure of the fit quality as in Sec. \ref{section:data_analysis} and \cite{Custodio_prl_2025}. It is interesting to note that this prediction is compatible with the data, suggesting that there is no particular anomaly in the deuteron multiplicity with respect to a purely statistical analysis. Moreover, the deviation of the predicted proton mass fraction from the experimental sample is in the direction of an overestimation of the data, while the opposite would be expected in the case of a spurious proton contribution coming from the decay of a deuteron into a proton and a neutron.
If final state deuteron break-up was present in the data, we would have a theoretical underestimation of the proton yield, which is at variance with our observation.
However, large uncertainties are observed.

\subsection{Bayesian Inference Results}

\begin{figure*}
	\centering
	\begin{subfigure}[b]{0.45\linewidth}		\includegraphics[width=\linewidth,height=7.5cm]{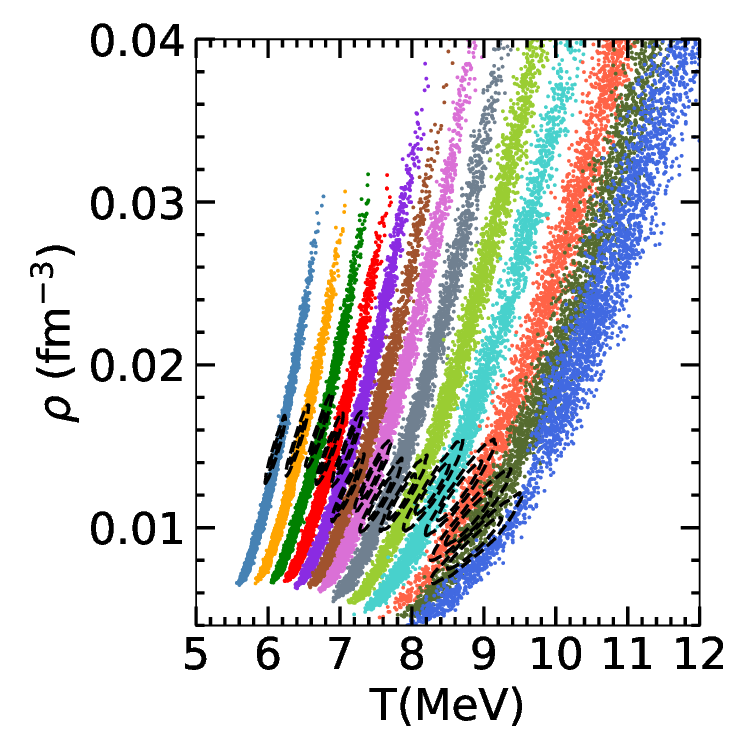}
    \end{subfigure}	
    \begin{subfigure}[b]{0.45\linewidth}
\includegraphics[width=\linewidth,height=7.5cm]{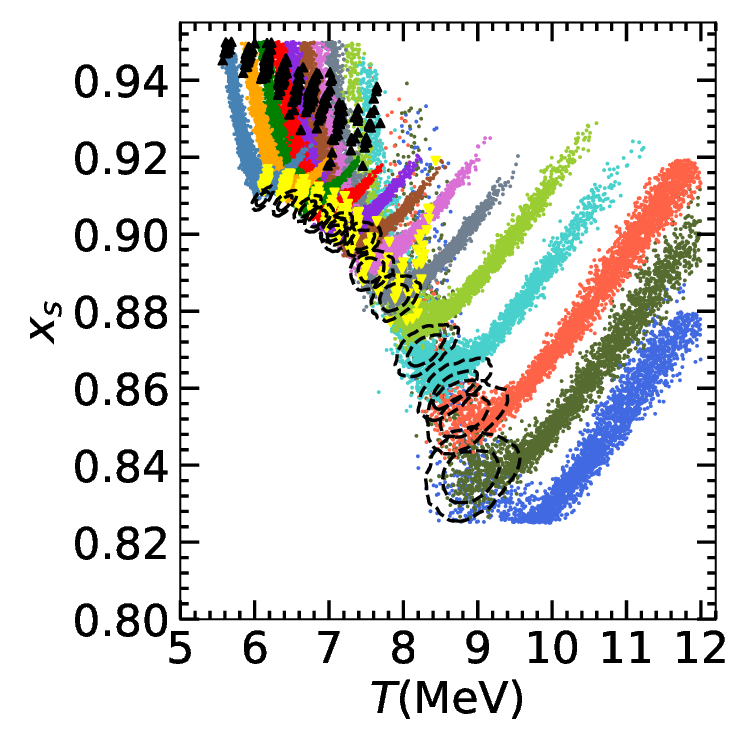}
	\end{subfigure}
 	\caption{\justifying { Each color represents a different $v_{\text{surf}}$(cm/ns) bin ranging from 4.1 (bottom) to 6.5 cm/ns (top) and each dot corresponds to a particular sample of the posterior distribution obtained after performing the Bayesian inference without the deuteron data. The unfilled black contours correspond to the 2-$\sigma$ posterior distributions obtained in Ref.\cite{Custodio_prl_2025} for the FSU RMF model. \textbf{Left: } Bayesian estimation of the thermodynamical parameters $\rho$ and $T$ for the colliding system $^{124}$Xe$+^{112}$Sn. 
   \textbf{Right: } Posterior distributions of the effective cluster coupling parameter $x_s$ as a function of the temperature $T$. The black triangles correspond to the solid black line in Fig.\ref{fig:2h_mass_fractions} and the inverted yellow triangles to the yellow solid line.
  }
  }
    \label{fig:T_vs_rho_and_xs_vs_T}
\end{figure*}

The inferred results of the temperature and the baryonic density are reported in the left panel of Fig.\ref{fig:T_vs_rho_and_xs_vs_T}. The results excluding the deuteron information are compatible with the findings of \cite{Custodio_prl_2025} (black dashed curves), but an important degeneracy is observed between temperature and density for each experimental sample.
Interestingly, the correlation between density and temperature is very steep, particularly for the lowest $v_{surf}$ bins. This explains why the full inference yields have larger uncertainties on the density estimations than on the temperature estimations.

Correspondingly, a complex pattern is observed in the posterior distribution of the parameter $x_s$, shown as a function of the estimated temperature in the right panel of Fig.\ref{fig:T_vs_rho_and_xs_vs_T}. Surprisingly, the results of the complete inference including the deuteron data as input in the Bayesian analysis (black dashed curves) almost exactly correspond to the minimal value of $x_s$ that is compatible with the  mass fractions {excluding the deuterons} in each $v_{\text{surf}}$ bin. This implies that, if the experimental deuteron information is not biased, the inclusion of the deuterons in the constraints is very important to pin down the correct value of $x_s$. This also means that, to keep compatibility with the (incomplete) set of data, any different estimation of temperature and density from the one of the complete fit should be accompanied by a smaller Pauli blocking effect{, i.e. a larger $x_s$}. This is qualitatively in agreement with the fact that in studies such as the one carried out by Pais et al. \cite{Pais2020prl}, where the same data set was considered but {only} the reduced information of the chemical equilibrium constants was included in the analysis, the global best fit value for the parameter $x_s$  was relatively high, $x_s = 0.92 \pm 0.02$. {Note that these $x_s$ values are compatible} {with} {the results obtained from the inference plotted in the right panel of Fig.\ref{fig:T_vs_rho_and_xs_vs_T}, for all $v_{\text{surf}}$ bins. }

\begin{figure*}
	\centering
		\includegraphics[width=0.7\linewidth]{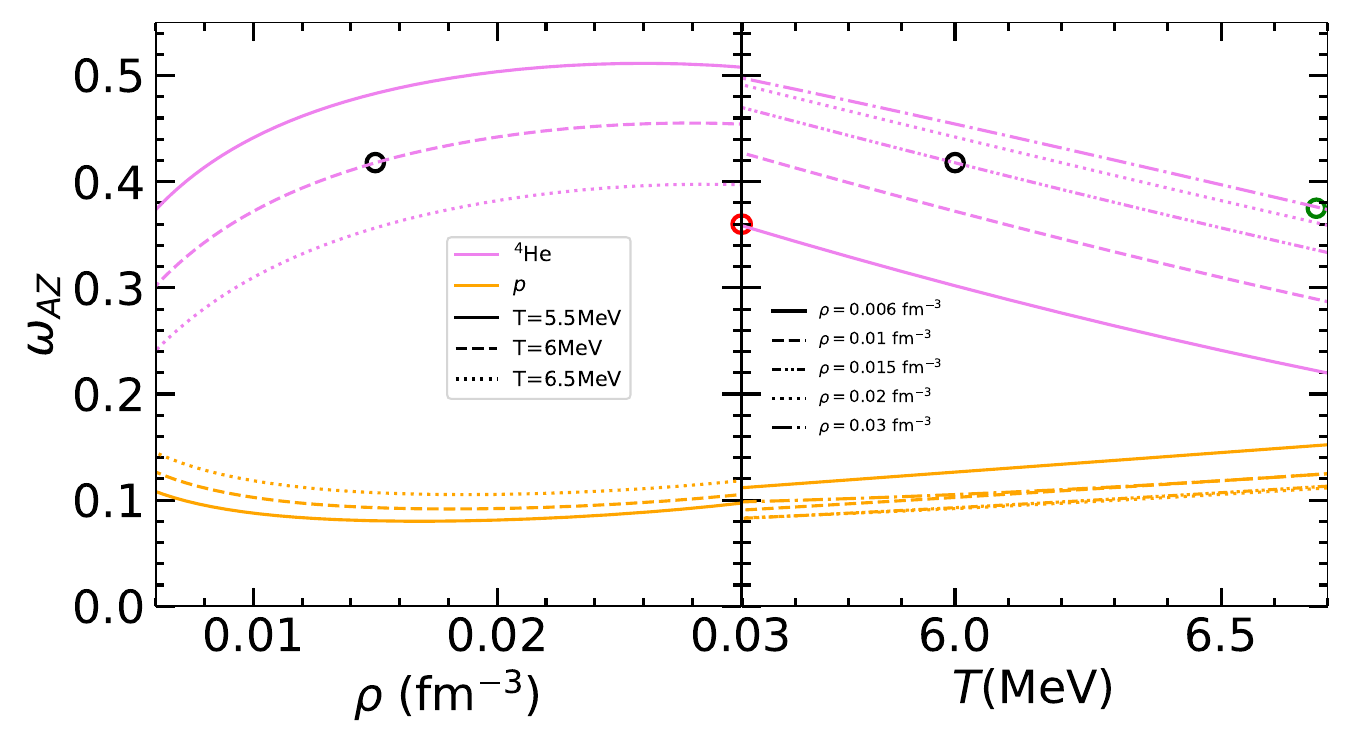}
	\caption{\justifying 
    {Theoretical prediction of proton (${}^{1}\text{H}$) and ${}^{4}\text{He}$ mass fractions for FSU RMF model taking as $x_s$ the temperature dependent $x_s(T)$ fit performed in Ref.\cite{Custodio_prl_2025}. The range of $\rho$ and $T$ displayed here correspond to those attained by the $v_{\text{surf}}=4.1$cm/ns bin (see Fig.\ref{fig:T_vs_rho_and_xs_vs_T}). Black circle represents the ${}^{4}\text{He}$ mass fraction for the lower $v_{\text{surf}}$ bin in Ref.\cite{Custodio_prl_2025}.
    \textbf{Left: } Mass fractions as a function of $\rho$ for fixed values of $T$. \textbf{Right: }Mass fractions as a function of $T$ for fixed values of $\rho$. Red circle the ${}^{4}\text{He}$ mass fraction for the lowest values of $T,\rho$ without considering the deuteron data in the inference; and the green circle the ${}^{4}\text{He}$ mass fraction for the highest values of $T,\rho$ without considering the deuteron data in the inference.
    }	
    }
	\label{fig:mass_fractions_fsu}
\end{figure*}

To better understand the non-monotonic behavior of the $x_s$ posterior distribution, we show in Fig.\ref{fig:mass_fractions_fsu} the theoretical behavior of the proton (${}^{1}\text{H}$) and $\alpha$ particle (${}^{4}\text{He}$) mass fractions at a fixed $y_p=0.45$, taken equal to the one of the entrance channel,
and using a temperature dependent $x_s(T)$ given by the quadratic fit of Ref.\cite{Custodio_prl_2025}. The results are shown for  the range of densities and temperatures obtained for the lowest $v_{\text{surf}}=4.1$cm$/$ns bin in the inference without the deuteron data. 
We can see that the most important information about the thermodynamical parameters is brought by ${}^{4}\text{He}$, while protons exhibit a weak dependence on density and temperature.

The ${}^{4}\text{He}$ mass fraction increases both by increasing the density at fixed temperature or by decreasing the temperature at fixed density, as expected.
In Ref.\cite{Custodio_prl_2025}, for the lowest $v_{\text{surf}}$ bin, the optimal $T$, $\rho$, and $x_s$ values are obtained at $T \approx 6\,\text{MeV}, \rho \approx 0.015 \,\text{fm$^{-3}$}, x_s \approx 0.91$. The corresponding ${}^{4}\text{He}$ mass fraction is signaled on  Fig.\ref{fig:mass_fractions_fsu} with a black circle. 
If we compare this ${}^{4}\text{He}$ mass fraction with the one obtained by considering the lowest (highest) $[T,\, \rho]$ values optimized in the inference without the deuteron data (see the extreme points of the lowest band in the left panel of Fig.  \ref{fig:T_vs_rho_and_xs_vs_T}), represented by the red (green) circle in Fig.\ref{fig:mass_fractions_fsu}, we conclude that in both limiting cases the ${}^{4}\text{He}$ mass fraction is lower than in the optimal case calculated in Ref.\cite{Custodio_prl_2025} (black circle). Therefore, to reproduce the same ${}^{4}\text{He}$ mass fraction with different [$T$, $\rho$] values, a stronger attraction in the ${}^{4}\text{He}$ channel (i.e. a larger $x_s$) is needed, which explains why the $x_s$ of Ref.\cite{Custodio_prl_2025} is at the minimum in Fig.\ref{fig:T_vs_rho_and_xs_vs_T}.

\begin{figure*}
	\centering
	\begin{subfigure}[b]{0.45\linewidth}	
    \includegraphics[width=\linewidth,height=7.5cm]{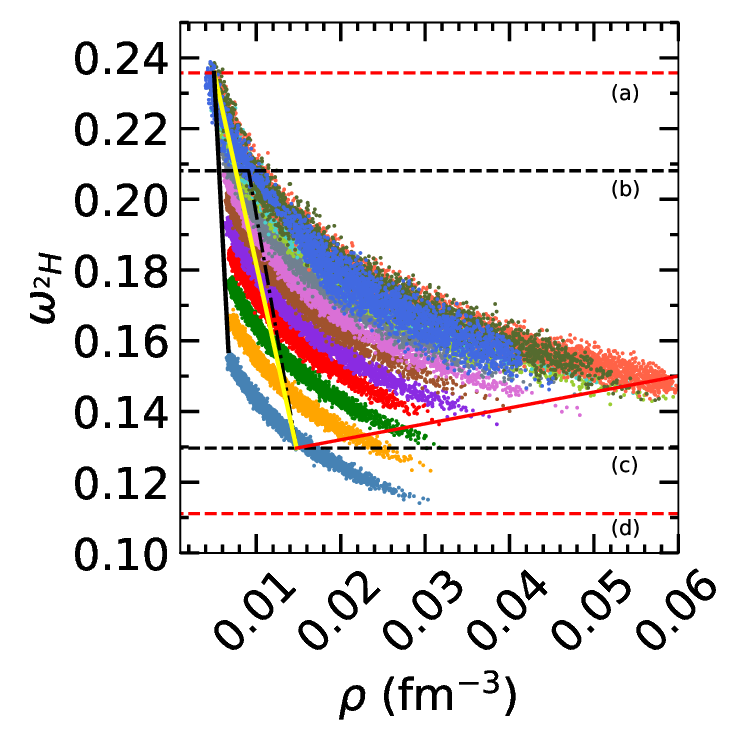}
    \end{subfigure}	
    \begin{subfigure}[b]{0.45\linewidth}
\includegraphics[width=\linewidth,height=7.5cm]{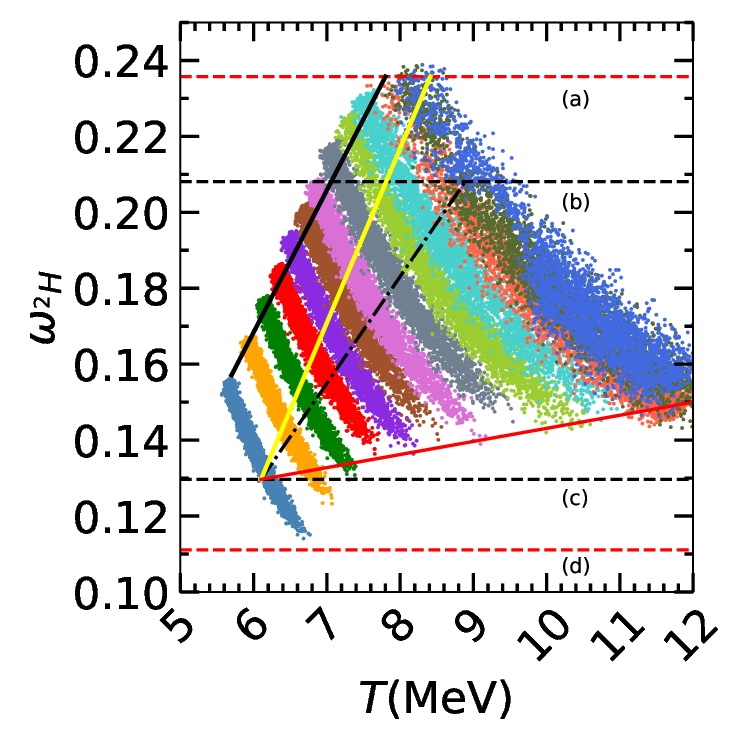}
	\end{subfigure}
 	\caption{ \justifying {Posterior distributions of the deuteron mass fractions as a function of $T$ (right side) and  $\rho$ (left side) calculated with the optimized ($T$, $\rho$, $x_s$) values without considering the deuteron data.  The horizontal dashed lines noted (a) and (d) correspond to the  maximum and minimum deuteron mass fraction predicted by the model, that are obtained respectively from the sample of highest (lowest) value of $v_{\text{surf}}$. The horizontal dashed lines
noted (b) and (c) give the maximum and minimum fraction in the experimental sample, also obtained respectively from the sample of highest (lowest) value of $v_{\text{surf}}$. The black dash-dotted line linearly joins the lowest and highest experimentally measured deuteron mass fraction; the black solid line corresponds to the qualitative interpolation of the lowest measured deuteron mass fraction to the highest predicted by the model; the yellow solid line corresponds to the qualitative interpolation between the maximum deuteron mass fractions for every $v_{\text{surf}}$; the red solid line corresponds to the qualitative interpolation between the experimental deuteron abundance at the lowest $v_{\text{surf}}$ and the lowest deuteron mass fraction (corresponding to the highest temperature value) at the highest $v_{\text{surf}}$. See text for details and explanations.
    }
    }
    \label{fig:2h_mass_fractions}
\end{figure*}

Generalizing the reasoning to the other $v_{\text{surf}}$ bins, we can say the following: to obtain a constant value of $\omega_{{}^{4}\text{He}}$ (which is the dominant cluster, and the one that shows the highest sensitivity to a temperature and density variation) with higher values of $T$ or lower values of $\rho$ with respect to the reference values $[T_{\text{ref}},\,\rho_{\text{ref}}]$ (here given by Ref.\cite{Custodio_prl_2025}), one needs a larger $x_s$ with respect to the reference $x_{s,\text{ref}}$. In our analysis of the ${}^{4}\text{He}$ mass fraction (Fig. \ref{fig:mass_fractions_fsu}), density and temperature are positively correlated,  the two effects going in opposite directions.
However, the optimal density $\rho_{\text{ref}}$ obtained in the analysis of Ref.\cite{Custodio_prl_2025} is close to the maximum cluster production density in the temperature domain covered by the experiment and above this density the production yields are relatively flat. For this reason, temperature plays a dominant role when we explore the right side of the $T,\,\rho$ correlation with respect to the reference $[T_{\text{ref}},\rho_{\text{ref}}]$ point {($T>T_{\text{ref}}$, $\rho>\rho_{\text{ref}}$)}, while density plays a dominant role when we explore the left side {($T<T_{\text{ref}}$, $\rho<\rho_{\text{ref}}$)}. This seems to be the reason for the peculiar correlation observed in the right part of Fig. \ref{fig:T_vs_rho_and_xs_vs_T}.

We now turn to examine the degeneracy in the thermodynamical parameter space observed in the left side of Fig.\ref{fig:T_vs_rho_and_xs_vs_T}.
This degeneracy is associated to the high uncertainty on the deuteron prediction shown by Fig.\ref{fig:Mass_Fractions_no_2h_in_likelihood_but_comparing_to_mass_fractions_with_2h}.  Fig.\ref{fig:2h_mass_fractions} shows the
posterior distributions of the deuteron fractions as a function of temperature and density, in the case where the deuterons are not used as a constraint in the analysis.
For each $v_{\text{surf}}$ bin, lower
(higher) temperatures and densities with respect to the estimation of Ref.\cite{Custodio_prl_2025}, displayed as black contours in Fig.\ref{fig:T_vs_rho_and_xs_vs_T} left, are compatible with the
data if we accept an overestimation (underestimation) of the deuteron fraction. 
In Fig.\ref{fig:2h_mass_fractions} (both left and right panels), the horizontal dashed lines denoted (a) and (d) give the maximum and minimum deuteron mass fraction predicted by
the model, which are obtained respectively from the sample of the highest and the lowest  value of $v_{\text{surf}}$. The horizontal dashed lines denoted (b) and (c) give the maximum and minimum  deuteron mass fractions in the experimental sample, also obtained, respectively, from the sample of the highest and the lowest value of $v_{\text{surf}}$.

The dashed-dotted lines in both left and right panels of Fig.\ref{fig:2h_mass_fractions} approximately join the deuteron mass fraction that is experimentally measured in each $v_{\text{surf}}$ bin. As discussed at length in Ref.\cite{Custodio_prl_2025}, this points to a relatively large temperature variation and
a relatively narrow density interval, a finding that was associated to the physical picture of a hot source cooling down in time, with a time dependence in the particle emission and particles that decouple at the surface of the source, that is at a relatively well defined (freeze-out) density.

In the hypothesis where the deuteron multiplicity would not be reliable because of an intrinsic deuteron 
\enquote{fragility}, the measured deuteron {mass} fraction should be taken as a lower limit to the actual deuteron abundance in equilibrium. If we still believe that the data excluding the deuterons can be analyzed with the statistical model, the maximum deuteron fraction is the one
approximately given by the black solid line in Fig.\ref{fig:2h_mass_fractions}, approximately joining the left side of the posterior distributions. This would lead to an approximately constant density $\rho \approx 0.005 \text{fm$^{-3}$}$  for the whole sample, slightly lower than the one extracted in Ref.\cite{Custodio_prl_2025}. The posterior distribution of the $x_s$ parameter corresponding to the black solid line in Fig.\ref{fig:2h_mass_fractions} is given by the black triangles in the left top corner of Fig.\ref{fig:T_vs_rho_and_xs_vs_T}. We still see a temperature dependence
of $x_s$, but the optimal values would be shifted with respect to the full inference of Ref.\cite{Custodio_prl_2025}.

However, the hypothesis of a maximal temperature independent deuteron break-up probability does not appear very realistic. 
A more realistic hypothesis about non-equilibrium effects in the deuteron behavior in dense matter could be that the deuteron breaking probability should increase for increasing temperature and/or density. A qualitative implementation of such condition is given by the yellow line in Fig.\ref{fig:2h_mass_fractions}, that corresponds to a linear interpolation between the deuteron mass fraction obtained in the experiment in the lowest $v_{\text{surf}}$ bin, and the maximum deuteron {mass} fraction compatible with the reproduction of the other multiplicities in the highest $v_{\text{surf}}$ bin. The corresponding $x_s$ posterior is given by the yellow triangles in Fig.\ref{fig:T_vs_rho_and_xs_vs_T}. We see that the estimation of the in-medium correction is perfectly compatible with the one of Ref.\cite{Custodio_prl_2025}. Concerning the trajectory in the ($T$, $\rho$) plane, with this hypothesis we partially recover 
a correlation between temperature and density. 
However, the density variation is very small and still consistent with the conclusions of Ref.\cite{Custodio_prl_2025}, namely the existence of a quasi-universal freeze-out density corresponding to the surface region of the participant region, where the clusters decouple from the cooling source.

Finally, we can see that the only possibility to reconcile the experimental results with the hypothesis of a significant increase in density with increasing temperature (i.e. an emission from an expanding cooling source without surface effects, as invoked in Qin et al. \cite{Qin2011}) is to  consider the possible path (shown in Fig.\ref{fig:2h_mass_fractions} by a red solid line)  of an increasing underestimation of deuterons with increasing $v_{\text{surf}}$. Such a path, starting from the experimental deuteron abundance at the lowest $v_{\text{surf}}$ and ending at the lowest deuteron mass fraction (corresponding to the highest temperature value) at the highest $v_{\text{surf}}$, corresponds to an approximately constant $x_s$ value, $x_s \approx 0.92$. This shows that the result of Pais et al. \cite{Pais2020prl} is compatible with the present analysis, and the difference with respect to Ref.\cite{Custodio_prl_2025} is essentially due to the reduction in information caused by the use of chemical constants instead of mass fractions for model calibration.
In light of the present analysis, we can also say that the solution of  Pais et al. \cite{Pais2020prl}  would correspond to an excess of deuteron production (deuteron \enquote{robustness} instead of deuteron \enquote{fragility}) in the experimental sample, due to the possible presence of a non-equilibrium early-time coalescence deuteron component that would increase the deuteron fraction, and correspondingly decrease the proton one.

\section{Conclusions} \label{section:conclusions}

The present paper is an extension and continuation of Ref.\cite{Custodio_prl_2025}.
The core motivation for that was to improve the poor description of experimental cluster abundances if the chemical equilibrium constants and the modified ideal gas approach developed in Refs.\cite{Pais2020,Pais2020prl} were considered to treat the  $^{136,124}$Xe$+^{124,112}$Sn central collisions at $32$MeV/nucleon data  measured with the INDRA apparatus \cite{Bougault_2018}. A good description of the particle fractions was achieved by performing an independent Bayesian analysis  on the experimentally measured mass fractions within the framework of a RMF model to describe the thermodynamic equilibrium of the samples \cite{Custodio_prl_2025}.
In the present work, the steps leading to the approach of Ref.\cite{Custodio_prl_2025} are discussed and the robustness of the conclusions with respect to hypotheses of the Bayesian scheme, the possible model dependence, and the possible presence of out-of-equilibrium effects in the experimental samples are analyzed in details.
In particular, different Bayesian schemes were explored to motivate and justify the posterior distributions of density and temperature obtained in Ref.\cite{Custodio_prl_2025}, and the degeneracy between a possible density and temperature dependence of the meson couplings are discussed. The important result of a single freeze-out density characterizing cluster formation in the experiment, is confirmed.

The possible degeneracy between $x_s$ and $x_{\omega}$ discussed in the literature \cite{Ferreira:2012ha} was also investigated. It was shown that, for the range of thermodynamical conditions explored by the data, it is equivalent to calibrate either ($\rho$,$T$,$x_s$) or ($\rho$,$T$,$x_{\omega}$) since both cases provide the same values for $\rho$ and $T$ as well as an excellent reproduction of experimental data. The only difference lies in the extracted values of $x_s$ and $x_{\omega}$, since the weakening attraction given by $x_s$ with temperature is replaced by an increasing repulsion provided by $x_{\omega}$ with temperature. Interestingly, 
the light cluster fractions obtained with the two different scenarios of cluster-meson couplings
are in reasonably good agreement even for values of density and temperature that are not directly constrained by the experimental data.
Small differences occur only for densities above the maximum fractions, with a $x_{\omega}(T,\rho)$ predicting slightly larger dissolution densities.

The impact of reducing the experimental information on the Bayesian inference of thermodynamic parameters was also studied. In particular, the  INDRA data corresponding to central collisions of the $^{124}$Xe$+^{112}$Sn system were reanalysed excluding from the Bayesian inference the deuteron multiplicity data. As a result, the unique solution obtained for each $v_{\text{surf}}$ sample in Ref.\cite{Custodio_prl_2025} evolves into a degenerate solution space corresponding to a positive correlation between temperature and density, and a greater indeterminacy in the $x_s$ ratio, especially at high $v_{\text{surf}}$.

This finding might explain why such a positive correlation was found in previous analyses \cite{Qin2011,Pais2020prl} that considered reduced information from the four equilibrium constants and a hypothesis on the source density. The prediction of the deuteron fraction obtained with this reduced information is well compatible with the data, suggesting that there is no a priori need to account for non-equilibrium effects or finite-state interactions. However, the prediction is not in contradiction with a possible overproduction or underproduction of deuteron in the experimental sample. If the latter hypothesis is retained, the resulting posterior distribution of the density and  $x_s$ parameter is still very well
compatible with the results of the complete inference, but the deduced temperature variation of the sample is slightly reduced. Finally, a constant value of $x_s \approx 0.92$ with temperature and density can only be retained if a non-negligible contribution of non-equilibrium deuterons from early stage coalescence processes contaminates the sample. 

However, the fact that the experimental deuteron fraction falls in the very middle of the model prediction 
suggests that the deuteron information can be considered statistical, and that deuterons can be safely added to the thermodynamic inference. 

Still, it would be extremely interesting to see if dynamical transport calculations with reliable production and destruction cross sections would support this interpretation.

\section*{Acknowledgements}

This work was partially supported by  the IN2P3 Master Project NewMAC, the ANR project `Gravitational waves from hot neutron stars and properties of ultra-dense matter' (GW-HNS, ANR-22-CE31-0001-01), national funds from FCT (Fundação para a Ciência e a Tecnologia, I.P, Portugal) for projects UID/04564/2025 identified by DOI 10.54499/UIDB/04564/2025 and 2024.16290.PEX with the associated DOI identifier 10.54499/2024.16290.PEX. T.C. acknowledges the grant PRT/BD/154193/2022
(FCT, Portugal) with DOI identifier https://doi.org/10.54499/PRT/BD/154193/2022. H.P. acknowledges the grant 2022.03966.CEECIND (FCT, Portugal) with DOI identifier 10.54499/2022.03966.CEECIND/CP1714/CT0004. The authors acknowledge the Laboratory for Advanced Computing at the University of Coimbra for providing {HPC} resources that have contributed to the research results reported within this paper, URL: \hyperlink{https://www.uc.pt/lca}{https://www.uc.pt/lca}. We acknowledge support from Région Normandie under RIN/FIDNEOS and
support from Portugal/France PESSOA program (PHC 47833UB France,
2021.09262.CBM Portugal).

\newpage 
\bibliographystyle{apsrev4-2}

\input{output.bbl}

\end{document}

%% file: output.bbl
%

%% file: submit.bbl
\begin{thebibliography}{34}%
\makeatletter
\providecommand \@ifxundefined [1]{%
 \@ifx{#1\undefined}
}%
\providecommand \@ifnum [1]{%
 \ifnum #1\expandafter \@firstoftwo
 \else \expandafter \@secondoftwo
 \fi
}%
\providecommand \@ifx [1]{%
 \ifx #1\expandafter \@firstoftwo
 \else \expandafter \@secondoftwo
 \fi
}%
\providecommand \natexlab [1]{#1}%
\providecommand \enquote  [1]{``#1''}%
\providecommand \bibnamefont  [1]{#1}%
\providecommand \bibfnamefont [1]{#1}%
\providecommand \citenamefont [1]{#1}%
\providecommand \href@noop [0]{\@secondoftwo}%
\providecommand \href [0]{\begingroup \@sanitize@url \@href}%
\providecommand \@href[1]{\@@startlink{#1}\@@href}%
\providecommand \@@href[1]{\endgroup#1\@@endlink}%
\providecommand \@sanitize@url [0]{\catcode `\\12\catcode `\$12\catcode
  `\&12\catcode `\#12\catcode `\^12\catcode `\_12\catcode `\%12\relax}%
\providecommand \@@startlink[1]{}%
\providecommand \@@endlink[0]{}%
\providecommand \url  [0]{\begingroup\@sanitize@url \@url }%
\providecommand \@url [1]{\endgroup\@href {#1}{\urlprefix }}%
\providecommand \urlprefix  [0]{URL }%
\providecommand \Eprint [0]{\href }%
\providecommand \doibase [0]{https://doi.org/}%
\providecommand \selectlanguage [0]{\@gobble}%
\providecommand \bibinfo  [0]{\@secondoftwo}%
\providecommand \bibfield  [0]{\@secondoftwo}%
\providecommand \translation [1]{[#1]}%
\providecommand \BibitemOpen [0]{}%
\providecommand \bibitemStop [0]{}%
\providecommand \bibitemNoStop [0]{.\EOS\space}%
\providecommand \EOS [0]{\spacefactor3000\relax}%
\providecommand \BibitemShut  [1]{\csname bibitem#1\endcsname}%
\let\auto@bib@innerbib\@empty
\bibitem [{\citenamefont {Arcones}\ \emph {et~al.}(2008)\citenamefont
  {Arcones}, \citenamefont {Mart\'{\i}nez-Pinedo}, \citenamefont {O'Connor},
  \citenamefont {Schwenk}, \citenamefont {Janka}, \citenamefont {Horowitz},\
  and\ \citenamefont {Langanke}}]{Arcones2008}%
  \BibitemOpen
  \bibfield  {author} {\bibinfo {author} {\bibfnamefont {A.}~\bibnamefont
  {Arcones}}, \bibinfo {author} {\bibfnamefont {G.}~\bibnamefont
  {Mart\'{\i}nez-Pinedo}}, \bibinfo {author} {\bibfnamefont {E.}~\bibnamefont
  {O'Connor}}, \bibinfo {author} {\bibfnamefont {A.}~\bibnamefont {Schwenk}},
  \bibinfo {author} {\bibfnamefont {H.-T.}\ \bibnamefont {Janka}}, \bibinfo
  {author} {\bibfnamefont {C.~J.}\ \bibnamefont {Horowitz}},\ and\ \bibinfo
  {author} {\bibfnamefont {K.}~\bibnamefont {Langanke}},\ }\href
  {https://doi.org/10.1103/PhysRevC.78.015806} {\bibfield  {journal} {\bibinfo
  {journal} {Phys. Rev. C}\ }\textbf {\bibinfo {volume} {78}},\ \bibinfo
  {pages} {015806} (\bibinfo {year} {2008})}\BibitemShut {NoStop}%
\bibitem [{\citenamefont {Sumiyoshi}\ and\ \citenamefont
  {Roepke}(2008)}]{Sumiyoshi:2008qv}%
  \BibitemOpen
  \bibfield  {author} {\bibinfo {author} {\bibfnamefont {K.}~\bibnamefont
  {Sumiyoshi}}\ and\ \bibinfo {author} {\bibfnamefont {G.}~\bibnamefont
  {Roepke}},\ }\href {https://doi.org/10.1103/PhysRevC.77.055804} {\bibfield
  {journal} {\bibinfo  {journal} {Phys. Rev. C}\ }\textbf {\bibinfo {volume}
  {77}},\ \bibinfo {pages} {055804} (\bibinfo {year} {2008})}\BibitemShut
  {NoStop}%
\bibitem [{\citenamefont {Furusawa}\ \emph {et~al.}(2013)\citenamefont
  {Furusawa}, \citenamefont {Nagakura}, \citenamefont {Sumiyoshi},\ and\
  \citenamefont {Yamada}}]{Furusawa:2013tta}%
  \BibitemOpen
  \bibfield  {author} {\bibinfo {author} {\bibfnamefont {S.}~\bibnamefont
  {Furusawa}}, \bibinfo {author} {\bibfnamefont {H.}~\bibnamefont {Nagakura}},
  \bibinfo {author} {\bibfnamefont {K.}~\bibnamefont {Sumiyoshi}},\ and\
  \bibinfo {author} {\bibfnamefont {S.}~\bibnamefont {Yamada}},\ }\href
  {https://doi.org/10.1088/0004-637X/774/1/78} {\bibfield  {journal} {\bibinfo
  {journal} {Astrophys. J.}\ }\textbf {\bibinfo {volume} {774}},\ \bibinfo
  {pages} {78} (\bibinfo {year} {2013})}\BibitemShut {NoStop}%
\bibitem [{\citenamefont {Bauswein}\ \emph {et~al.}(2013)\citenamefont
  {Bauswein}, \citenamefont {Goriely},\ and\ \citenamefont
  {Janka}}]{Bauswein:2013yna}%
  \BibitemOpen
  \bibfield  {author} {\bibinfo {author} {\bibfnamefont {A.}~\bibnamefont
  {Bauswein}}, \bibinfo {author} {\bibfnamefont {S.}~\bibnamefont {Goriely}},\
  and\ \bibinfo {author} {\bibfnamefont {H.~T.}\ \bibnamefont {Janka}},\ }\href
  {https://doi.org/10.1088/0004-637X/773/1/78} {\bibfield  {journal} {\bibinfo
  {journal} {Astrophys. J.}\ }\textbf {\bibinfo {volume} {773}},\ \bibinfo
  {pages} {78} (\bibinfo {year} {2013})}\BibitemShut {NoStop}%
\bibitem [{\citenamefont {Rosswog}(2015)}]{Rosswog2015}%
  \BibitemOpen
  \bibfield  {author} {\bibinfo {author} {\bibfnamefont {S.}~\bibnamefont
  {Rosswog}},\ }\href {https://doi.org/10.1142/S0218271815300128} {\bibfield
  {journal} {\bibinfo  {journal} {Int. J. Mod. Phys. D}\ }\textbf {\bibinfo
  {volume} {24}},\ \bibinfo {pages} {1530012} (\bibinfo {year}
  {2015})}\BibitemShut {NoStop}%
\bibitem [{\citenamefont {Radice}\ \emph {et~al.}(2018)\citenamefont {Radice},
  \citenamefont {Perego}, \citenamefont {Hotokezaka}, \citenamefont {Fromm},
  \citenamefont {Bernuzzi},\ and\ \citenamefont {Roberts}}]{Radice:2018pdn}%
  \BibitemOpen
  \bibfield  {author} {\bibinfo {author} {\bibfnamefont {D.}~\bibnamefont
  {Radice}}, \bibinfo {author} {\bibfnamefont {A.}~\bibnamefont {Perego}},
  \bibinfo {author} {\bibfnamefont {K.}~\bibnamefont {Hotokezaka}}, \bibinfo
  {author} {\bibfnamefont {S.~A.}\ \bibnamefont {Fromm}}, \bibinfo {author}
  {\bibfnamefont {S.}~\bibnamefont {Bernuzzi}},\ and\ \bibinfo {author}
  {\bibfnamefont {L.~F.}\ \bibnamefont {Roberts}},\ }\href
  {https://doi.org/10.3847/1538-4357/aaf054} {\bibfield  {journal} {\bibinfo
  {journal} {Astrophys. J.}\ }\textbf {\bibinfo {volume} {869}},\ \bibinfo
  {pages} {130} (\bibinfo {year} {2018})}\BibitemShut {NoStop}%
\bibitem [{\citenamefont {Nav\'o}\ \emph {et~al.}(2023)\citenamefont {Nav\'o},
  \citenamefont {Reichert}, \citenamefont {Obergaulinger},\ and\ \citenamefont
  {Arcones}}]{Navo:2022xle}%
  \BibitemOpen
  \bibfield  {author} {\bibinfo {author} {\bibfnamefont {G.}~\bibnamefont
  {Nav\'o}}, \bibinfo {author} {\bibfnamefont {M.}~\bibnamefont {Reichert}},
  \bibinfo {author} {\bibfnamefont {M.}~\bibnamefont {Obergaulinger}},\ and\
  \bibinfo {author} {\bibfnamefont {A.}~\bibnamefont {Arcones}},\ }\href
  {https://doi.org/10.3847/1538-4357/acd640} {\bibfield  {journal} {\bibinfo
  {journal} {Astrophys. J.}\ }\textbf {\bibinfo {volume} {951}},\ \bibinfo
  {pages} {112} (\bibinfo {year} {2023})}\BibitemShut {NoStop}%
\bibitem [{\citenamefont {Psaltis}\ \emph {et~al.}(2024)\citenamefont
  {Psaltis}, \citenamefont {Jacobi}, \citenamefont {Montes}, \citenamefont
  {Arcones}, \citenamefont {Hansen},\ and\ \citenamefont
  {Schatz}}]{Psaltis:2023jvk}%
  \BibitemOpen
  \bibfield  {author} {\bibinfo {author} {\bibfnamefont {A.}~\bibnamefont
  {Psaltis}}, \bibinfo {author} {\bibfnamefont {M.}~\bibnamefont {Jacobi}},
  \bibinfo {author} {\bibfnamefont {F.}~\bibnamefont {Montes}}, \bibinfo
  {author} {\bibfnamefont {A.}~\bibnamefont {Arcones}}, \bibinfo {author}
  {\bibfnamefont {C.~J.}\ \bibnamefont {Hansen}},\ and\ \bibinfo {author}
  {\bibfnamefont {H.}~\bibnamefont {Schatz}},\ }\href
  {https://doi.org/10.3847/1538-4357/ad2dfb} {\bibfield  {journal} {\bibinfo
  {journal} {Astrophys. J.}\ }\textbf {\bibinfo {volume} {966}},\ \bibinfo
  {pages} {11} (\bibinfo {year} {2024})}\BibitemShut {NoStop}%
\bibitem [{\citenamefont {Oertel}\ \emph {et~al.}(2017)\citenamefont {Oertel}
  \emph {et~al.}}]{Oertel2017}%
  \BibitemOpen
  \bibfield  {author} {\bibinfo {author} {\bibfnamefont {M.}~\bibnamefont
  {Oertel}} \emph {et~al.},\ }\href@noop {} {\bibfield  {journal} {\bibinfo
  {journal} {Rev. Mod. Phys.}\ }\textbf {\bibinfo {volume} {89}},\ \bibinfo
  {pages} {015007} (\bibinfo {year} {2017})}\BibitemShut {NoStop}%
\bibitem [{\citenamefont {Ropke}(2011)}]{Ropke:2011tr}%
  \BibitemOpen
  \bibfield  {author} {\bibinfo {author} {\bibfnamefont {G.}~\bibnamefont
  {Ropke}},\ }\href {https://doi.org/10.1016/j.nuclphysa.2011.07.010}
  {\bibfield  {journal} {\bibinfo  {journal} {Nucl. Phys. A}\ }\textbf
  {\bibinfo {volume} {867}},\ \bibinfo {pages} {66} (\bibinfo {year} {2011})},\
  \Eprint {https://arxiv.org/abs/1101.4685} {arXiv:1101.4685 [nucl-th]}
  \BibitemShut {NoStop}%
\bibitem [{\citenamefont {R\"opke}(2015)}]{Ropke2015}%
  \BibitemOpen
  \bibfield  {author} {\bibinfo {author} {\bibfnamefont {G.}~\bibnamefont
  {R\"opke}},\ }\href {https://doi.org/10.1103/PhysRevC.92.054001} {\bibfield
  {journal} {\bibinfo  {journal} {Phys. Rev. C}\ }\textbf {\bibinfo {volume}
  {92}},\ \bibinfo {pages} {054001} (\bibinfo {year} {2015})},\ \Eprint
  {https://arxiv.org/abs/1411.4593} {arXiv:1411.4593 [nucl-th]} \BibitemShut
  {NoStop}%
\bibitem [{\citenamefont {R\"opke}(2020)}]{Ropke_2020}%
  \BibitemOpen
  \bibfield  {author} {\bibinfo {author} {\bibfnamefont {G.}~\bibnamefont
  {R\"opke}},\ }\href {https://doi.org/10.1103/PhysRevC.101.064310} {\bibfield
  {journal} {\bibinfo  {journal} {Phys. Rev. C}\ }\textbf {\bibinfo {volume}
  {101}},\ \bibinfo {pages} {064310} (\bibinfo {year} {2020})}\BibitemShut
  {NoStop}%
\bibitem [{\citenamefont {Ren}\ \emph {et~al.}(2024)\citenamefont {Ren},
  \citenamefont {Elhatisari}, \citenamefont {Lähde}, \citenamefont {Lee},\
  and\ \citenamefont {Meißner}}]{Ren_2024}%
  \BibitemOpen
  \bibfield  {author} {\bibinfo {author} {\bibfnamefont {Z.}~\bibnamefont
  {Ren}}, \bibinfo {author} {\bibfnamefont {S.}~\bibnamefont {Elhatisari}},
  \bibinfo {author} {\bibfnamefont {T.~A.}\ \bibnamefont {Lähde}}, \bibinfo
  {author} {\bibfnamefont {D.}~\bibnamefont {Lee}},\ and\ \bibinfo {author}
  {\bibfnamefont {U.-G.}\ \bibnamefont {Meißner}},\ }\href
  {https://doi.org/https://doi.org/10.1016/j.physletb.2024.138463} {\bibfield
  {journal} {\bibinfo  {journal} {Physics Letters B}\ }\textbf {\bibinfo
  {volume} {850}},\ \bibinfo {pages} {138463} (\bibinfo {year}
  {2024})}\BibitemShut {NoStop}%
\bibitem [{\citenamefont {Typel}\ \emph {et~al.}(2015)\citenamefont {Typel},
  \citenamefont {Oertel},\ and\ \citenamefont {Kl\"ahn}}]{CompOse}%
  \BibitemOpen
  \bibfield  {author} {\bibinfo {author} {\bibfnamefont {S.}~\bibnamefont
  {Typel}}, \bibinfo {author} {\bibfnamefont {M.}~\bibnamefont {Oertel}},\ and\
  \bibinfo {author} {\bibfnamefont {T.}~\bibnamefont {Kl\"ahn}},\ }\href
  {https://doi.org/10.1134/S1063779615040061} {\bibfield  {journal} {\bibinfo
  {journal} {Phys. Part. Nucl.}\ }\textbf {\bibinfo {volume} {46}},\ \bibinfo
  {pages} {633} (\bibinfo {year} {2015})},\ \Eprint
  {https://arxiv.org/abs/1307.5715} {arXiv:1307.5715 [astro-ph.SR]}
  \BibitemShut {NoStop}%
\bibitem [{\citenamefont {Typel}\ \emph {et~al.}(2010)\citenamefont {Typel},
  \citenamefont {Ropke}, \citenamefont {Klahn}, \citenamefont {Blaschke},\ and\
  \citenamefont {Wolter}}]{Typel2009}%
  \BibitemOpen
  \bibfield  {author} {\bibinfo {author} {\bibfnamefont {S.}~\bibnamefont
  {Typel}}, \bibinfo {author} {\bibfnamefont {G.}~\bibnamefont {Ropke}},
  \bibinfo {author} {\bibfnamefont {T.}~\bibnamefont {Klahn}}, \bibinfo
  {author} {\bibfnamefont {D.}~\bibnamefont {Blaschke}},\ and\ \bibinfo
  {author} {\bibfnamefont {H.~H.}\ \bibnamefont {Wolter}},\ }\href
  {https://doi.org/10.1103/PhysRevC.81.015803} {\bibfield  {journal} {\bibinfo
  {journal} {Phys. Rev. C}\ }\textbf {\bibinfo {volume} {81}},\ \bibinfo
  {pages} {015803} (\bibinfo {year} {2010})},\ \Eprint
  {https://arxiv.org/abs/0908.2344} {arXiv:0908.2344 [nucl-th]} \BibitemShut
  {NoStop}%
\bibitem [{\citenamefont {Pais}\ \emph {et~al.}(2019)\citenamefont {Pais},
  \citenamefont {Gulminelli}, \citenamefont {Provid\^encia},\ and\
  \citenamefont {R\"opke}}]{Pais_2019}%
  \BibitemOpen
  \bibfield  {author} {\bibinfo {author} {\bibfnamefont {H.}~\bibnamefont
  {Pais}}, \bibinfo {author} {\bibfnamefont {F.}~\bibnamefont {Gulminelli}},
  \bibinfo {author} {\bibfnamefont {C.}~\bibnamefont {Provid\^encia}},\ and\
  \bibinfo {author} {\bibfnamefont {G.}~\bibnamefont {R\"opke}},\ }\href
  {https://doi.org/10.1103/PhysRevC.99.055806} {\bibfield  {journal} {\bibinfo
  {journal} {Phys. Rev. C}\ }\textbf {\bibinfo {volume} {99}},\ \bibinfo
  {pages} {055806} (\bibinfo {year} {2019})}\BibitemShut {NoStop}%
\bibitem [{\citenamefont {Qin}\ \emph {et~al.}(2012)\citenamefont {Qin} \emph
  {et~al.}}]{Qin2011}%
  \BibitemOpen
  \bibfield  {author} {\bibinfo {author} {\bibfnamefont {L.}~\bibnamefont
  {Qin}} \emph {et~al.},\ }\href
  {https://doi.org/10.1103/PhysRevLett.108.172701} {\bibfield  {journal}
  {\bibinfo  {journal} {Phys. Rev. Lett.}\ }\textbf {\bibinfo {volume} {108}},\
  \bibinfo {pages} {172701} (\bibinfo {year} {2012})},\ \Eprint
  {https://arxiv.org/abs/1110.3345} {arXiv:1110.3345 [nucl-ex]} \BibitemShut
  {NoStop}%
\bibitem [{\citenamefont {Pais}\ \emph
  {et~al.}(2020{\natexlab{a}})\citenamefont {Pais} \emph {et~al.}}]{Pais2020}%
  \BibitemOpen
  \bibfield  {author} {\bibinfo {author} {\bibfnamefont {H.}~\bibnamefont
  {Pais}} \emph {et~al.},\ }\href {https://doi.org/10.1088/1361-6471/aba561}
  {\bibfield  {journal} {\bibinfo  {journal} {J. Phys. G}\ }\textbf {\bibinfo
  {volume} {47}},\ \bibinfo {pages} {105204} (\bibinfo {year}
  {2020}{\natexlab{a}})},\ \Eprint {https://arxiv.org/abs/2006.07256}
  {arXiv:2006.07256 [nucl-th]} \BibitemShut {NoStop}%
\bibitem [{\citenamefont {Pais}\ \emph
  {et~al.}(2020{\natexlab{b}})\citenamefont {Pais} \emph
  {et~al.}}]{Pais2020prl}%
  \BibitemOpen
  \bibfield  {author} {\bibinfo {author} {\bibfnamefont {H.}~\bibnamefont
  {Pais}} \emph {et~al.},\ }\href
  {https://doi.org/10.1103/PhysRevLett.125.012701} {\bibfield  {journal}
  {\bibinfo  {journal} {Phys. Rev. Lett.}\ }\textbf {\bibinfo {volume} {125}},\
  \bibinfo {pages} {012701} (\bibinfo {year} {2020}{\natexlab{b}})}\BibitemShut
  {NoStop}%
\bibitem [{\citenamefont {Cust\'odio}\ \emph {et~al.}(2025)\citenamefont
  {Cust\'odio}, \citenamefont {Rebillard-Souli\'e}, \citenamefont {Bougault},
  \citenamefont {Gruyer}, \citenamefont {Gulminelli}, \citenamefont {Malik},
  \citenamefont {Pais},\ and\ \citenamefont
  {Provid\^encia}}]{Custodio_prl_2025}%
  \BibitemOpen
  \bibfield  {author} {\bibinfo {author} {\bibfnamefont {T.}~\bibnamefont
  {Cust\'odio}}, \bibinfo {author} {\bibfnamefont {A.}~\bibnamefont
  {Rebillard-Souli\'e}}, \bibinfo {author} {\bibfnamefont {R.}~\bibnamefont
  {Bougault}}, \bibinfo {author} {\bibfnamefont {D.}~\bibnamefont {Gruyer}},
  \bibinfo {author} {\bibfnamefont {F.}~\bibnamefont {Gulminelli}}, \bibinfo
  {author} {\bibfnamefont {T.}~\bibnamefont {Malik}}, \bibinfo {author}
  {\bibfnamefont {H.}~\bibnamefont {Pais}},\ and\ \bibinfo {author}
  {\bibfnamefont {C.}~\bibnamefont {Provid\^encia}},\ }\href
  {https://doi.org/10.1103/PhysRevLett.134.082304} {\bibfield  {journal}
  {\bibinfo  {journal} {Phys. Rev. Lett.}\ }\textbf {\bibinfo {volume} {134}},\
  \bibinfo {pages} {082304} (\bibinfo {year} {2025})}\BibitemShut {NoStop}%
\bibitem [{\citenamefont {Ferreira}\ and\ \citenamefont
  {Providencia}(2012)}]{Ferreira:2012ha}%
  \BibitemOpen
  \bibfield  {author} {\bibinfo {author} {\bibfnamefont {M.}~\bibnamefont
  {Ferreira}}\ and\ \bibinfo {author} {\bibfnamefont {C.}~\bibnamefont
  {Providencia}},\ }\href {https://doi.org/10.1103/PhysRevC.85.055811}
  {\bibfield  {journal} {\bibinfo  {journal} {Phys. Rev. C}\ }\textbf {\bibinfo
  {volume} {85}},\ \bibinfo {pages} {055811} (\bibinfo {year} {2012})},\
  \Eprint {https://arxiv.org/abs/1206.0139} {arXiv:1206.0139 [nucl-th]}
  \BibitemShut {NoStop}%
\bibitem [{\citenamefont {Pais}\ \emph {et~al.}(2015)\citenamefont {Pais},
  \citenamefont {Chiacchiera},\ and\ \citenamefont
  {Provid\^encia}}]{Pais:2015xoa}%
  \BibitemOpen
  \bibfield  {author} {\bibinfo {author} {\bibfnamefont {H.}~\bibnamefont
  {Pais}}, \bibinfo {author} {\bibfnamefont {S.}~\bibnamefont {Chiacchiera}},\
  and\ \bibinfo {author} {\bibfnamefont {C.}~\bibnamefont {Provid\^encia}},\
  }\href {https://doi.org/10.1103/PhysRevC.91.055801} {\bibfield  {journal}
  {\bibinfo  {journal} {Phys. Rev. C}\ }\textbf {\bibinfo {volume} {91}},\
  \bibinfo {pages} {055801} (\bibinfo {year} {2015})},\ \Eprint
  {https://arxiv.org/abs/1504.03964} {arXiv:1504.03964 [nucl-th]} \BibitemShut
  {NoStop}%
\bibitem [{\citenamefont {Pais}\ \emph {et~al.}(2018)\citenamefont {Pais},
  \citenamefont {Gulminelli}, \citenamefont {Provid\^encia},\ and\
  \citenamefont {R\"opke}}]{Pais2018}%
  \BibitemOpen
  \bibfield  {author} {\bibinfo {author} {\bibfnamefont {H.}~\bibnamefont
  {Pais}}, \bibinfo {author} {\bibfnamefont {F.}~\bibnamefont {Gulminelli}},
  \bibinfo {author} {\bibfnamefont {C.}~\bibnamefont {Provid\^encia}},\ and\
  \bibinfo {author} {\bibfnamefont {G.}~\bibnamefont {R\"opke}},\ }\href
  {https://doi.org/10.1103/PhysRevC.97.045805} {\bibfield  {journal} {\bibinfo
  {journal} {Phys. Rev. C}\ }\textbf {\bibinfo {volume} {97}},\ \bibinfo
  {pages} {045805} (\bibinfo {year} {2018})}\BibitemShut {NoStop}%
\bibitem [{\citenamefont {Todd-Rutel}\ and\ \citenamefont
  {Piekarewicz}(2005)}]{Todd-Rutel:2005yzo}%
  \BibitemOpen
  \bibfield  {author} {\bibinfo {author} {\bibfnamefont {B.~G.}\ \bibnamefont
  {Todd-Rutel}}\ and\ \bibinfo {author} {\bibfnamefont {J.}~\bibnamefont
  {Piekarewicz}},\ }\href {https://doi.org/10.1103/PhysRevLett.95.122501}
  {\bibfield  {journal} {\bibinfo  {journal} {Phys. Rev. Lett.}\ }\textbf
  {\bibinfo {volume} {95}},\ \bibinfo {pages} {122501} (\bibinfo {year}
  {2005})},\ \Eprint {https://arxiv.org/abs/nucl-th/0504034}
  {arXiv:nucl-th/0504034} \BibitemShut {NoStop}%
\bibitem [{\citenamefont {Cust\'odio}\ \emph {et~al.}(2020)\citenamefont
  {Cust\'odio} \emph {et~al.}}]{Custodio2020}%
  \BibitemOpen
  \bibfield  {author} {\bibinfo {author} {\bibfnamefont {T.}~\bibnamefont
  {Cust\'odio}} \emph {et~al.},\ }\href
  {https://doi.org/10.1140/epja/s10050-020-00302-w} {\bibfield  {journal}
  {\bibinfo  {journal} {Eur. Phys. J. A}\ }\textbf {\bibinfo {volume} {56}},\
  \bibinfo {pages} {295} (\bibinfo {year} {2020})},\ \Eprint
  {https://arxiv.org/abs/2009.14035} {arXiv:2009.14035 [nucl-th]} \BibitemShut
  {NoStop}%
\bibitem [{\citenamefont {Bougault}\ \emph {et~al.}(2018)\citenamefont
  {Bougault} \emph {et~al.}}]{Bougault_2018}%
  \BibitemOpen
  \bibfield  {author} {\bibinfo {author} {\bibfnamefont {R.}~\bibnamefont
  {Bougault}} \emph {et~al.} (\bibinfo {collaboration} {INDRA Collaboration}),\
  }\href {https://doi.org/10.1103/PhysRevC.97.024612} {\bibfield  {journal}
  {\bibinfo  {journal} {Phys. Rev. C}\ }\textbf {\bibinfo {volume} {97}},\
  \bibinfo {pages} {024612} (\bibinfo {year} {2018})}\BibitemShut {NoStop}%
\bibitem [{\citenamefont {Bougault}\ \emph {et~al.}(2020)\citenamefont
  {Bougault} \emph {et~al.}}]{indra}%
  \BibitemOpen
  \bibfield  {author} {\bibinfo {author} {\bibfnamefont {R.}~\bibnamefont
  {Bougault}} \emph {et~al.},\ }\href
  {https://doi.org/10.1088/1361-6471/ab56ba} {\bibfield  {journal} {\bibinfo
  {journal} {J. Phys. G}\ }\textbf {\bibinfo {volume} {47}},\ \bibinfo {pages}
  {025103} (\bibinfo {year} {2020})},\ \Eprint
  {https://arxiv.org/abs/1911.08355} {arXiv:1911.08355 [nucl-ex]} \BibitemShut
  {NoStop}%
\bibitem [{\citenamefont {Rebillard-Souli{\'e}}\ \emph
  {et~al.}(2024)\citenamefont {Rebillard-Souli{\'e}} \emph
  {et~al.}}]{Rebillard-Soulie:2023rqu}%
  \BibitemOpen
  \bibfield  {author} {\bibinfo {author} {\bibfnamefont {A.}~\bibnamefont
  {Rebillard-Souli{\'e}}} \emph {et~al.},\ }\href
  {https://doi.org/10.1088/1361-6471/ad0edd} {\bibfield  {journal} {\bibinfo
  {journal} {J. Phys. G}\ }\textbf {\bibinfo {volume} {51}},\ \bibinfo {pages}
  {015104} (\bibinfo {year} {2024})},\ \Eprint
  {https://arxiv.org/abs/2311.05392} {arXiv:2311.05392 [nucl-ex]} \BibitemShut
  {NoStop}%
\bibitem [{\citenamefont {Gelman}\ \emph {et~al.}(2013)\citenamefont {Gelman},
  \citenamefont {Carlin}, \citenamefont {Stern}, \citenamefont {Dunson},
  \citenamefont {Vehtari},\ and\ \citenamefont {Rubin}}]{Gelman2013}%
  \BibitemOpen
  \bibfield  {author} {\bibinfo {author} {\bibfnamefont {A.}~\bibnamefont
  {Gelman}}, \bibinfo {author} {\bibfnamefont {J.~B.}\ \bibnamefont {Carlin}},
  \bibinfo {author} {\bibfnamefont {H.~S.}\ \bibnamefont {Stern}}, \bibinfo
  {author} {\bibfnamefont {D.~B.}\ \bibnamefont {Dunson}}, \bibinfo {author}
  {\bibfnamefont {A.}~\bibnamefont {Vehtari}},\ and\ \bibinfo {author}
  {\bibfnamefont {D.~B.}\ \bibnamefont {Rubin}},\ }\href@noop {} {\emph
  {\bibinfo {title} {Bayesian Data Analysis}}},\ \bibinfo {edition} {3rd}\ ed.\
  (\bibinfo  {publisher} {CRC Press},\ \bibinfo {year} {2013})\BibitemShut
  {NoStop}%
\bibitem [{\citenamefont {Buchner}\ \emph {et~al.}(2014)\citenamefont
  {Buchner}, \citenamefont {Georgakakis}, \citenamefont {Nandra}, \citenamefont
  {Hsu}, \citenamefont {Rangel}, \citenamefont {Brightman}, \citenamefont
  {Merloni}, \citenamefont {Salvato}, \citenamefont {Donley},\ and\
  \citenamefont {Kocevski}}]{Buchner:2014nha}%
  \BibitemOpen
  \bibfield  {author} {\bibinfo {author} {\bibfnamefont {J.}~\bibnamefont
  {Buchner}}, \bibinfo {author} {\bibfnamefont {A.}~\bibnamefont
  {Georgakakis}}, \bibinfo {author} {\bibfnamefont {K.}~\bibnamefont {Nandra}},
  \bibinfo {author} {\bibfnamefont {L.}~\bibnamefont {Hsu}}, \bibinfo {author}
  {\bibfnamefont {C.}~\bibnamefont {Rangel}}, \bibinfo {author} {\bibfnamefont
  {M.}~\bibnamefont {Brightman}}, \bibinfo {author} {\bibfnamefont
  {A.}~\bibnamefont {Merloni}}, \bibinfo {author} {\bibfnamefont
  {M.}~\bibnamefont {Salvato}}, \bibinfo {author} {\bibfnamefont
  {J.}~\bibnamefont {Donley}},\ and\ \bibinfo {author} {\bibfnamefont
  {D.}~\bibnamefont {Kocevski}},\ }\href
  {https://doi.org/10.1051/0004-6361/201322971} {\bibfield  {journal} {\bibinfo
   {journal} {Astron. Astrophys.}\ }\textbf {\bibinfo {volume} {564}},\
  \bibinfo {pages} {A125} (\bibinfo {year} {2014})},\ \Eprint
  {https://arxiv.org/abs/1402.0004} {arXiv:1402.0004 [astro-ph.HE]}
  \BibitemShut {NoStop}%
\bibitem [{\citenamefont {Buchner}(2023)}]{Buchner:2021kpm}%
  \BibitemOpen
  \bibfield  {author} {\bibinfo {author} {\bibfnamefont {J.}~\bibnamefont
  {Buchner}},\ }\href {https://doi.org/10.1214/23-SS144} {\bibfield  {journal}
  {\bibinfo  {journal} {Statistics Surveys}\ }\textbf {\bibinfo {volume}
  {17}},\ \bibinfo {pages} {169 } (\bibinfo {year} {2023})}\BibitemShut
  {NoStop}%
\bibitem [{\citenamefont {Skilling}(2004)}]{Skilling2004}%
  \BibitemOpen
  \bibfield  {author} {\bibinfo {author} {\bibfnamefont {J.}~\bibnamefont
  {Skilling}},\ }\href {https://doi.org/10.1063/1.1835238} {\bibfield
  {journal} {\bibinfo  {journal} {AIP Conference Proceedings}\ }\textbf
  {\bibinfo {volume} {735}},\ \bibinfo {pages} {395} (\bibinfo {year}
  {2004})},\ \Eprint
  {https://arxiv.org/abs/https://aip.scitation.org/doi/pdf/10.1063/1.1835238}
  {https://aip.scitation.org/doi/pdf/10.1063/1.1835238} \BibitemShut {NoStop}%
\bibitem [{\citenamefont {Rebillard-Soulie}(2024)}]{Alex_Thesis_2024}%
  \BibitemOpen
  \bibfield  {author} {\bibinfo {author} {\bibfnamefont {A.}~\bibnamefont
  {Rebillard-Soulie}},\ }\href {https://theses.hal.science/tel-04901901}
  {\bibinfo {type} {Theses}},\ \bibinfo  {school} {{Normandie Universit{\'e}}}
  (\bibinfo {year} {2024})\BibitemShut {NoStop}%
\bibitem [{\citenamefont {Albergo}\ \emph {et~al.}(1985)\citenamefont {Albergo}
  \emph {et~al.}}]{albergo}%
  \BibitemOpen
  \bibfield  {author} {\bibinfo {author} {\bibfnamefont {S.}~\bibnamefont
  {Albergo}} \emph {et~al.},\ }\bibfield  {journal} {\bibinfo  {journal} {Il
  Nuovo Cimento A}\ }\textbf {\bibinfo {volume} {89}},\ \href
  {https://doi.org/https://doi.org/10.1007/BF02773614}
  {https://doi.org/10.1007/BF02773614} (\bibinfo {year} {1985})\BibitemShut
  {NoStop}%
\end{thebibliography}
